\documentclass[11pt]{article}
\pdfoutput=1
\usepackage{graphicx}
\usepackage{hyperref}
\usepackage{epsfig}
\usepackage{amsmath}
\usepackage{amsthm}
\usepackage{amssymb}
\usepackage{multirow}
\usepackage{color}
\usepackage[nosort]{cite}
\usepackage{hyperref}
\usepackage[fontsize=11.45pt]{scrextend}
\usepackage{braket,mathrsfs,slashed}
\usepackage{float}
\usepackage{setspace}
\usepackage[caption=false]{subfig}
\newlength{\abstractwidth}
\newcommand{\be}{\begin{equation}}
\newcommand{\ee}{\end{equation}}

\renewcommand{\title}[1]{\vbox{\center\bf{\Large{#1}}}\vspace{5mm}}
\renewcommand{\author}[1]{\vbox{\center#1}\vspace{5mm}}
\newcommand{\address}[1]{\vbox{\center\em#1}}
\newcommand{\email}[1]{\vbox{\center\tt#1}\vspace{5mm}}

\usepackage{dsfont}

\usepackage[utf8]{inputenc}
\usepackage{rotating}
\usepackage{tikz-feynman}

\usepackage{dsfont}
\usepackage[paper=a4paper,margin=1.5cm]{geometry}
\usepackage[utf8]{inputenc}
\usepackage{rotating}
\usepackage{soul}

\usepackage{mathrsfs} 
\usepackage{bbm}
\usepackage{etoolbox} 
\usepackage{multirow}

\renewcommand\[{\begin{equation}}
\renewcommand\]{\end{equation}}

\newcommand{\ba}{\begin{eqnarray}}
\newcommand{\ea}{\end{eqnarray}}

\hypersetup{pdfstartview=FitV,colorlinks=true,linkcolor=midblue,citecolor=midblue,filecolor=midblue,urlcolor=midblue}

\definecolor{midblue}{rgb}{0,0,0.5}

\begin{document}
	
		\newgeometry{top=3.1cm,bottom=3.1cm,right=2.4cm,left=2.4cm}
		
	\begin{titlepage}
	\begin{center}
		\hfill \\
		\vskip 0.5cm

		\title{Massless and Partially Massless Limits\\[2mm] in Quadratic Gravity}

		\author{\large Luca Buoninfante}
		
		\address{Nordita,
			Stockholm University and KTH Royal Institute of Technology,\\
			Hannes Alfvéns väg 12, SE-106 91 Stockholm, Sweden}
		\email{\rm \href{mailto:luca.buoninfante@su.se}{luca.buoninfante@su.se}}

	\end{center}

\begin{abstract}
In the context of perturbative quantum field theory, the addition of quadratic-curvature invariants to the Einstein-Hilbert action makes it possible to achieve strict renormalizability in four dimensions. The additional terms $R^2$ and $C_{\mu\nu\rho\sigma} C^{\mu\nu\rho\sigma}$ are multiplied by dimensionless coefficients that are related to the masses of the extra gravitational degrees of freedom and to the interaction couplings. The aim of this paper is to study the limit of the theory in which the Weyl-squared coefficient tends to infinity. Remarkably, the result of this limit turns out to be sensitive to the presence of a cosmological constant:  when the latter is zero we have a massless limit for the spin-2 ghost, while when the cosmological constant is different from zero we obtain a partially massless limit. We show that the renormalizability property and the  ghost-like nature of the massive spin-$2$ field  ensure that the two limits do not hit strong couplings, unlike standard ghost-free theories of massive gravity. In particular, in the partially massless limit the interactions mediated by the spin-$2$ sector vanish. We argue that our results can be useful for understanding the high-energy limit of Quadratic Gravity.
\end{abstract}

\end{titlepage}

{
	\hypersetup{linkcolor=black}
	\tableofcontents
}

\baselineskip=17.63pt



\newpage

\section{Introduction and motivations}

The criterion of renormalizability is a powerful guiding principle to select fundamental theories in the framework of perturbative Quantum Field Theory (QFT)~\cite{Weinberg:1995mt}. 
Indeed, in the Standard Model of particle physics the electromagnetic, weak and strong interactions are described by strictly renormalizable QFTs.

One may wonder if the same QFT paradigm can be employed to understand quantum aspects of the gravitational interaction at a fundamental level. On the one hand, it is notoriously known that Einstein's general relativity is perturbatively non-renormalizable~\cite{tHooft:1974toh,Goroff:1985sz,vandeVen:1991gw} since the interaction couplings are given by negative powers of the Planck mass. On the other hand, it is very natural to ask whether we can still find a renormalizable theory of quantum gravity that extends Einstein's general relativity in the ultraviolet (UV) regime. The answer to this question is yes.

In fact, in four spacetime dimensions there exists a \textit{unique}\footnote{Super-renormalizable higher-curvature models have also been considered in the literature but they are not unique~\cite{Anselmi:2017ygm} (see also~\cite{Buoninfante:2022ykf} for a review on higher-derivative models). Indeed, unlike super-renormalizability, the property of \textit{strict} renormalizability imposes stronger constraints on the allowed terms in a local Lagrangian and, therefore, is more restrictive for the selection of fundamental  theories.} local and strictly renormalizable QFT of gravity that is diffeomorphism invariant, metric compatible, and that recovers Einstein's theory in the low-energy regime. It is known as Quadratic Gravity and  its action contains all possible operators up to dimension four in units of mass~\cite{Stelle:1976gc}:
\begin{equation}
	S[g]=\frac{1}{2}\int {\rm d}^4x\sqrt{-g}\,\left[M_{\rm p}^2 (R-2\Lambda)+\frac{\alpha}{6}R^2-\frac{\beta}{2}C_{\mu\nu\rho\sigma}C^{\mu\nu\rho\sigma}  \right]\,,\label{action-stelle}
\end{equation}
where $R=g^{\mu\nu}R_{\mu\nu}$ and $C_{\mu\nu\rho\sigma}$ are the Ricci scalar and Weyl tensor, respectively; we did not write down the invariant $R_{\mu\nu}R^{\mu\nu}$ or $R_{\mu\nu\rho\sigma}R^{\mu\nu\rho\sigma}$ because in four dimensions they can be eliminated via the Gauss-Bonnet identity. Note that, despite being a total derivative for spacetimes topologically equivalent to Minkowski in four dimensions, the Gauss-Bonnet term ${\rm GB}=R_{\mu\nu\rho\sigma}R^{\mu\nu\rho\sigma}-4R_{\mu\nu}R^{\mu\nu}+R^2$ is important for the renormalization of the theory, as is $\Box R$. Given our convention for the choice of parameters in eq.~\eqref{action-stelle}, $M_{\rm p}\sim 10^{18}$GeV is the reduced Planck mass, $\Lambda$ the cosmological constant with  mass-squared dimension, while $\alpha$ and $\beta$ are two dimensionless coefficients.  
The quadratic-curvature invariants introduce new massive degrees of freedom: a spin-$0$ coming from $R^2$ and a spin-$2$ from $C_{\mu\nu\rho\sigma}C^{\mu\nu\rho\sigma},$ whose presence ameliorates the high-energy behavior of the theory, e.g., the propagator falls off  like $\sim 1/p^4$ in the UV regime.

Remarkably, the criterion of strict renormalizability naturally provides a cosmological constant term and an extra massive scalar field whose self-potential happens to have a very suitable shape for describing the inflationary phase in the early Universe~\cite{Starobinsky:1980te,Ivanov:2016hcm,Salvio:2017xul,Anselmi:2020lpp}. $\Lambda$ and $\alpha$ are free parameters whose renormalized physical values can be fixed with experiments. Indeed, according to the $\Lambda$CDM scenario~\cite{SupernovaSearchTeam:1998fmf} the late time accelerated expansion of our Universe can be described by fixing the cosmological constant to be of order
$\Lambda\simeq  10^{-122}M_{\rm p}^2$~\cite{Planck:2018vyg}, while CMB observations can be explained by setting $\alpha \simeq 10^{10}$ at inflationary energy scales. In Quadratic Gravity the primordial scalar fluctuations can be described by Starobinsky inflation~\cite{Starobinsky:1980te,Vilenkin:1985md} which, contrary to the usual beliefs, it is not just a model but can be consistently embedded in a renormalizable theory of quantum gravity.

Despite the improved UV behavior and the outstanding achievements in explaining cosmological observations, the price to pay for having renormalizability is that the additional spin-$2$ field turns out to be a \textit{ghost}~\cite{Stelle:1976gc}, which in general can cause Ostrogradsky instabilities~\cite{Woodard:2015zca} and  break unitarity when standard quantization prescriptions are implemented.

This unusual feature of the theory has motivated recent investigations towards understanding the spin-$2$ ghost in Quadratic Gravity. Among the most popular proposals, there are quantizations assuming negative norms for ghost-like states~\cite{Salvio:2018crh,Holdom:2021hlo,Holdom:2021oii,Holdom:2023usn} (see also~\cite{Woodard:2023tgb} for a critical review and~\cite{Strumia:2017dvt,Salvio:2020axm,Salvio:2019wcp} for possible physical interpretations of the negative norms), or quantizations implementing an alternative prescription for shifting the poles in the propagator such that the ghost is converted into a purely virtual particle which does not belong to the set of on-shell states~\cite{Anselmi:2017ygm,Anselmi:2018ibi,Anselmi:2018tmf,Piva:2023bcf}. Another proposal is to treat the ghost as an unstable particle propagating positive energy backward in time~\cite{Donoghue:2019fcb,Donoghue:2019ecz}.
Strictly speaking, all these approaches are formulated around the Minkowski spacetime and address the question of stability  mainly at the quantum level; however, see~\cite{Salvio:2017xul,Anselmi:2020lpp} for some extrapolations to cosmological backgrounds. 

A ghost-like spin-2 also appears in the context of conformal gravity~\cite{Rachwal:2022pfe} where approaches to remove the ghost were proposed both at the quantum~\cite{Mannheim:2011ds} and the classical~\cite{Maldacena:2011mk,Hell:2023rbf} levels. However, it is important to remark that conformal gravity, i.e., pure Weyl-squared gravity, is \textit{not} renormalizable beyond one loop because the $R^2$ counterterm is generated at two loops~\cite{Fradkin:1983tg}.

\subsubsection*{Aim of this work}

In the present paper we aim to investigate some new aspects of Quadratic Gravity that may provide new insights into the dynamics of the spin-$2$ ghost beyond Minkowski spacetime and at both the classical and quantum levels, including its high-energy behavior, and regardless of the specific quantization procedure chosen for the ghost. 

We are interested in studying an extreme regime in which we expect the Weyl-squared term to be non-negligible and the ghost dynamics to become important. In particular, we want to ask what is the limit of the action~\eqref{action-stelle} in which the Weyl-squared coefficient tends to infinity, i.e.,
\begin{equation}
\lim\limits_{\beta\rightarrow \infty}S=\,?\label{limit-beta}
\end{equation}
While this question may seem like just a mathematical curiosity, its answer proves to be very useful for understanding new physical features of the spin-$2$ ghost. As we will explain below, the limit $\beta\rightarrow \infty$ can be equivalently rephrased as a limit for the mass of the spin-$2$ ghost. This aspect allows us to make a comparison between Quadratic Gravity and standard ghost-free theories of massive gravity: we will learn that the ghost-like nature of the massive spin-$2$ field together with the property of renormalizability prevent strong couplings and the need of screening mechanisms that are typically invoked in theories of massive gravity~\cite{Hinterbichler:2011tt,deRham:2014zqa}.

We will clarify that a clean procedure to take the limit~\eqref{limit-beta} is to keep the canonically normalized fields fixed. This is similar to what happens in Yang-Mills theory where the coupling constant $g$ can either multiply the entire Lagrangian, i.e., $\frac{1}{g^2}\hat{F}_{\mu\nu}\hat{F}^{\mu\nu}$ where $\hat{F}_{\mu\nu}=\partial_\mu\hat{A}_{\nu}-\partial_\nu\hat{A}_\mu,$ or can be moved to the interaction vertices by rescaling the vector field as $A_\mu= \frac{1}{g} \hat{A}_\mu$ and $F_{\mu\nu}=\frac{1}{g}\hat{F}_{\mu\nu}$.  The Lagrangian written in terms of the canonically normalized field $A_\mu$ is more convenient from a perturbative QFT perspective because the limit of zero interaction coupling, i.e., $g\rightarrow 0$ with $A_\mu\,=\,\text{fixed}$, would consistently give a free Lagrangian. 
Therefore, a convenient way to perform our analysis is to first recast eq.~\eqref{action-stelle} into an equivalent canonical form in which only second order derivatives appear and in such a way that the kinetic and interaction terms of all gravitational degrees of freedom (massless spin-$2$, massive spin-$2$ and massive spin-$0$) are explicitly identified. 
We will notice that the mass of the spin-$2$ ghost and all interaction couplings depend non-trivially on the cosmological constant.

Let us explain the organization of this paper and briefly summarize the content of each section.
\begin{description}
	
	\item[Sec.~\ref{sec:action}:] We rewrite the action~\eqref{action-stelle} in the second-order canonical form in which the contributions of the canonically normalized spin-$0$ and spin-$2$ massive fields are made explicit. This task will be achieved through the help of field transformations. The new form of the action will be very useful to identify the masses of the additional degrees of freedom and the various interaction couplings independently of the background metric. 
	
	\item[Sec.~\ref{sec:decouplings}:] We analyse the limit $\beta\rightarrow \infty$  of the theory~\eqref{action-stelle} and show that its meaning highly depends on the presence of the cosmological constant. In particular, we note that $\beta\rightarrow \infty$ corresponds to a \textit{massless limit} ($m_2^2\rightarrow 0$) for the spin-$2$ ghost mass when $\Lambda=0$ and to a \textit{partially massless limit} ($m_2^2\rightarrow \frac{2}{3}\Lambda$)~\cite{Deser:1983mm,Deser:1983tm,Deser:2001us,Hassan:2012gz,Hassan:2013pca,deRham:2013wv,Joung:2014aba,DeRham:2018axr} when $\Lambda> 0.$ This implies that the way in which the five helicities of the spin-$2$ ghost are organized in the limit $\beta\rightarrow \infty$  highly depends on the presence of a cosmological constant. To correctly take into account all degrees of freedom we employ the St\"uckelberg formalism. When $\Lambda=0$ we find that all the degrees of freedom are interacting; while, when $\Lambda> 0$ the spin-$2$ sector (massless graviton $+$ partially massless graviton) becomes free and completely decouples.

	\item[Sec.~\ref{sec:phys-impl}:] We argue that the results obtained in sec.~\ref{sec:decouplings} can be useful for gaining a new understanding of the high-energy limit of Quadratic Gravity, in particular in the case of a non-zero and positive cosmological constant. We first remind that $\alpha$ and $\beta$ run with the energy~\cite{Julve:1978xn,Fradkin:1981iu,Avramidi:1985ki,Salvio:2014soa,Anselmi:2018ibi}: in particular, when they are both positive $\beta$ grows and $\alpha$ decreases as the energy increases. Thus, we provide arguments suggesting that  the resulting theory in the limit $\beta\rightarrow \infty$ with $\Lambda>0$ can describe Quadratic Gravity~\eqref{action-stelle} in the deep UV and discuss some implications.

	\item[Sec.~\ref{sec:conclus}:] It is devoted to summary of the main results and  concluding remarks. In particular, we will comment on the role the cosmological constant in Quadratic Gravity.

\end{description}

Throughout the entire work we always adopt the Natural Units system $\hbar=1=c$ and the mostly positive convention for the metric signature ($-+++$). We choose the following convention for the Riemann and Ricci tensors: $[\nabla_\nu,\nabla_\rho]V^\sigma=V^\mu R^\sigma_{\mu\nu\rho},$  $[\nabla_\nu,\nabla_\rho]V_\mu=-V_\sigma R^\sigma_{\mu\nu\rho},$  $R^{\sigma}_{\mu\nu\rho}=\partial_\nu\Gamma^\sigma_{\mu\rho}-\partial_\rho\Gamma^\sigma_{\mu\nu}+\Gamma^\sigma_{\alpha\nu}\Gamma^\alpha_{\mu\rho}-\Gamma^\sigma_{\alpha\rho}\Gamma_{\mu\nu}^\alpha,$  and $R_{\mu\rho}=R^\nu_{\mu\nu\rho}=\delta^{\nu}_\sigma R^{\sigma}_{\mu\nu\rho}=g^{\sigma\nu}R_{\sigma\mu\nu\rho}.$ Moreover, we only work with a non-negative cosmological constant:  $\Lambda=0$ or $\Lambda>0.$

\section{Action in canonical form}\label{sec:action}

In this section we recast the action~\eqref{action-stelle} into an equivalent second order form by explicitly separating the contributions of the massive spin-$0$ and spin-$2$ fields with the help of metric transformations and field redefinitions. To do so, we implement the strategy followed in Refs.~\cite{Anselmi:2018tmf,Kaku:1977pa,Hindawi:1995an,Hinterbichler:2015soa,Kubo:2022jwu} and, in addition, we will make important remarks on the canonical normalization of the fields and highlight the peculiar role played by the cosmological constant.

\subsection{Massive spin-$0$}\label{sec:spin-0}

To separate the contribution of the massive spin-$0$ we introduce an auxiliary scalar field $\hat{\omega}$ such that we can rewrite~\eqref{action-stelle} as
\begin{equation}
S[g,\hat{\omega}]=S_{\rm EH}[g]+S_{\rm W}[g]+\frac{\alpha}{12}\int {\rm d}^4x\sqrt{-g} \left[a\left(-R+4\Lambda\right)\hat{\omega}+b\,\hat{\omega}^2\right]+A[g]\,,\label{0-aux-action-1}
\end{equation}
where 
\begin{equation}
S_{\rm EH}[g]= \frac{M_{\rm p}^2}{2}\int{\rm d}^4x\sqrt{-g}(R-2\Lambda)\label{EH-action}
\end{equation}
is the Einstein-Hilbert action and
\begin{equation}
	S_{\rm W}[g]= -\frac{\beta}{4}\int{\rm d}^4x\sqrt{-g}C_{\mu\nu\rho\sigma}C^{\mu\nu\rho\sigma}\,,\label{W-action}
\end{equation}
the Weyl-squared term. The constants $a,$ $b$ and the functional $A[g]$ can be determined by imposing that the expression in eq.~\eqref{action-stelle} is recovered upon using the field equation $\delta S/\delta \hat{\omega}=0$ $\Leftrightarrow$ $\hat{\omega}=\frac{a}{2b}(R-4\Lambda).$ By doing so, we get
\begin{equation}
\frac{a^2}{4b}=-1\,,\qquad A[g]=\frac{4\alpha\Lambda}{3}\frac{1}{2}\int{\rm d}^4x\sqrt{-g}(R-2\Lambda)\,.\label{unkown-a-A}
\end{equation}
Since we are interested in diagonalizing the kinetic operator, without any loss of generality it is convenient to choose $a=2$ and $b=-1$, thus eq.~\eqref{0-aux-action-1} becomes
\begin{equation}
	S[g,\hat{\omega}]=\frac{\bar{M}_{\rm p}^2}{M_{\rm p}^2}S_{\rm EH}[g]+S_{\rm W}[g]+\frac{\alpha}{12}\int {\rm d}^4x\sqrt{-g} \left(2R-8\Lambda-\hat{\omega}\right)\hat{\omega}\,,\label{0-aux-action-2}
\end{equation}
where we have defined a shifted Planck mass,
\begin{equation}
\bar{M}_{\rm p}^2\equiv M_{\rm p}^2+\frac{4}{3}\Lambda\alpha\,.\label{gamma-bar}
\end{equation}

To  diagonalize the kinetic operator for the metric $g_{\mu\nu}$ and the scalar field $\hat{\omega}$ and eliminate terms linear in $\hat{\omega},$ we perform the Weyl transformation
\begin{equation}
g_{\mu\nu}\rightarrow \left(\frac{1}{1+\alpha\,\hat{\omega}/3\bar{M}_{\rm p}^2}\right)g_{\mu\nu}\,,\label{conformal-transf}
\end{equation}
that is well-defined as long as $1+\alpha\hat{\omega}/3\bar{M}^2_{\rm p}>0,$ which is always the case in the context of perturbative QFT and more generally for sub-Planckian field configurations. After some algebra we obtain
\begin{equation}
S[g,\hat{\omega}]=\frac{\bar{M}_{\rm p}^2}{M_{\rm p}^2}S_{\rm EH}[g]+S_{\rm W}[g]+S_0[g,\hat{\omega}]\,, \label{omega-action-compl}
\end{equation}
where
\begin{equation}
	S_{0}[g,\hat{\omega}]= \frac{\alpha^2}{6\bar{M}_{\rm p}^2}\int {\rm d}^4x\sqrt{-g}\frac{1}{\left(1+\alpha\hat{\omega}/3\bar{M}_{\rm p}^2\right)^2}\left(-\frac{1}{2}\nabla^\mu\hat{\omega}\nabla_\mu \hat{\omega}-\frac{m_0^2}{2}\hat{\omega}^2\right)\,,\label{omega-action-non-pert}
\end{equation}
and we have defined the mass of the scalar field 
\begin{equation}
	m_0^2\equiv \frac{M_{\rm p}^2}{\alpha}\,, \label{phi-mass}
\end{equation}
which can be identified by expanding around $\hat{\omega}=0$ and looking at the term quadratic in $\hat{\omega}.$

\paragraph{Fixing signs.} The overall coefficient in~\eqref{omega-action-non-pert} has to be reabsorbed through a field redefinition in order to obtain a canonical form. To do so, we first have to fix its sign. This can be done by requiring that the spin-$0$ degree of freedom is not a ghost, i.e.,
\begin{equation}
\bar{M}_{\rm p}^2=M_{\rm p}^2 +\frac{4}{3}\Lambda\alpha>0\,.\label{gamma-bar>0}
\end{equation}
In addition, to avoid the spin-$0$ to be a tachyon we must impose $m_0^2\geq 0$ which implies  
\begin{equation}
\alpha>0\,, \label{alpha>0}
\end{equation}
where we have obviously assumed that $M_{\rm p}^2>0.$ In this paper we only work with a non-negative cosmological constant, i.e. $\Lambda \geq 0,$ therefore the no-tachyon condition~\eqref{alpha>0} also ensures~\eqref{gamma-bar>0}.

\paragraph{Canonical normalization.} Having fixed the overall sign in the action~\eqref{omega-action-non-pert}, we can now canonically normalize the scalar field as
\begin{equation}
	\omega= \sqrt{\frac{3}{2}}\frac{\alpha}{3}\frac{\hat{\omega}}{\bar{M}_{\rm p}}\,.\label{field-chi}
\end{equation}
Then, if we make the field redefinition
\begin{equation}
	\phi=-\sqrt{\frac{3}{2}}\bar{M}_{\rm p}\log\left(1+\sqrt{\frac{2}{3}}\frac{\omega}{\bar{M}_{\rm p}}\right)\,,
	\label{tilde-phi-def}
\end{equation}
we can recast~\eqref{omega-action-non-pert} in the following canonical form:
\begin{equation}
S_0[g,\phi]=\int {\rm d}^4x\sqrt{-g}\left[ -\frac{1}{2}\nabla_\mu\phi\nabla^\mu \phi-\frac{m_0^2}{2}\frac{3\bar{M}^2_{\rm p}}{2}\left(1-e^{\sqrt{\frac{2}{3}}\phi/\bar{M}_{\rm p}}\right)^2  \right]
\end{equation}

An important point to emphasize is that when $\Lambda\neq 0$  we need to shift the Planck mass in the Einstein-Hilbert action, i.e., $M_{\rm p}^2 \rightarrow \bar{M}_{\rm p}^2,$ in order to diagonalize the kinetic operator for $g_{\mu\nu}$ and $\phi$. This feature makes the interaction couplings dependent on the cosmological constant. An analogue property holds for the massive spin-$2$ field as we will now show.

\subsection{Massive spin-$2$}\label{sec:spin-2}

In four dimensions we can use the identity $C_{\mu\nu\rho\sigma}C^{\mu\nu\rho\sigma}=R_{\mu\nu\rho\sigma}R^{\mu\nu\rho\sigma}-2R_{\mu\nu}R^{\mu\nu}+\frac{2}{3}R^2$ and the fact that the topological Gauss-Bonnet invariant ${\rm GB}=R_{\mu\nu\rho\sigma}R^{\mu\nu\rho\sigma}-4R_{\mu\nu}R^{\mu\nu}+R^2$ is a total derivative to rewrite the Weyl-squared term in the action as
\begin{equation}
	S[g,\phi]=\frac{\bar{M}_{\rm p}^2}{M_{\rm p}^2}S_{\rm EH}[g]+ S_{0}[g,\phi]-\frac{\beta}{2}\int {\rm d}^4x\sqrt{-g}\left(R_{\mu\nu}R^{\mu\nu}-\frac{1}{3}R^2\right)\,.\label{action-weyl->Ricci}
\end{equation}
To separate the contribution of the massive spin-$2$ we  introduce an auxiliary two-rank symmetric field $\hat{f}_{\mu\nu}$ such that we can write
\begin{equation}
	S[g,\phi,\hat{f}]=\frac{\bar{M}_{\rm p}^2}{M_{\rm p}^2}S_{\rm EH}[g]+ S_{0}[g,\phi]+B[g]	-\frac{\beta}{2}\int \!{\rm d}^4x\sqrt{-g}\left[c\left(G_{\mu\nu}+\Lambda g_{\mu\nu}\right)\hat{f}^{\mu\nu}+d\left(\hat{f}^{\mu\nu}\hat{f}_{\mu\nu}-\hat{f}^2\right)\right]\,,
	\label{aux-spin-2-B}
\end{equation}
where $G_{\mu\nu}=R_{\mu\nu}-\frac{1}{2}g_{\mu\nu}R$ is the Einstein tensor and $\hat{f}=g^{\mu\nu}\hat{f}_{\mu\nu}$ denotes the trace. The constants $c,$ $d$ and the functional $B[g]$ can be determined by imposing that the expression in eq.~\eqref{action-weyl->Ricci} is recovered upon using the field equations $\delta S/\delta \hat{f}^{\mu\nu}=0$ $\Leftrightarrow$ $\hat{f}_{\mu\nu}=-\frac{c}{2d}(R_{\mu\nu}-\frac{1}{6}g_{\mu\nu}R-\frac{1}{3}\Lambda g_{\mu\nu}).$ By doing so, we obtain
\begin{equation}
	\frac{c^2}{4d}=-1\,,\qquad B[g]=\frac{2\beta\Lambda}{3}\frac{1}{2}\int{\rm d}^4x\sqrt{-g}(R-2\Lambda)\,.\label{unkown-spin-2}
\end{equation}
Then, by choosing $c=\bar{M}_{\rm p}^2 $ and $d=-\bar{M}_{\rm p}^4/4$ without any loss of generality and using  
\begin{equation}
\frac{M_{\rm p}^2}{2}\sqrt{-g}\left(G^{\mu\nu}+\Lambda g^{\mu\nu}\right)=-\frac{\delta S_{\rm EH}}{\delta g_{\mu\nu}}\,,
\end{equation}
we can rewrite the action as
\begin{equation}
	S[g,\phi]=\frac{\tilde{M}_{\rm p}^2}{M_{\rm p}^2}S_{\rm EH}[g]+ S_{0}[g,\phi]	+\beta\int {\rm d}^4x \left[\frac{\bar{M}_{\rm p}^2}{M_{\rm p}^2}\frac{\delta S_{\rm EH}}{\delta g_{\mu\nu}}\hat{f}_{\mu\nu}+\sqrt{-g}\frac{\bar{M}_{\rm p}^4}{8}\left(\hat{f}^{\mu\nu}\hat{f}_{\mu\nu}-\hat{f}^2\right)\right]\,,\label{action-aux-spin-2}
\end{equation}
where we have defined a doubly shifted Planck mass,
\begin{equation}
\tilde{M}_{\rm p}^2\equiv \bar{M}_{\rm p}^2+\frac{2}{3}\beta\Lambda=M_{\rm p}^2+\frac{2}{3}(2\alpha+\beta)\Lambda\,.
\end{equation}

The next task is to diagonalize the kinetic operator for the metric tensor $g_{\mu\nu}$ and the massive spin-$2$ field $\hat{f}_{\mu\nu}$. This can be done by shifting the metric as
\begin{equation}
g_{\mu\nu}\rightarrow g_{\mu\nu}-\beta \left(\frac{\bar{M}_{\rm p}}{\tilde{M}_{\rm p}}\right)^2\hat{f}_{\mu\nu}\,,\label{metric-shift}
\end{equation}
thus expanding the action~\eqref{action-aux-spin-2} perturbatively in $\hat{f}_{\mu\nu}$ the term $\frac{\delta S_{\rm EH}}{\delta g_{\mu\nu}}\hat{f}_{\mu\nu}$  cancels out and  we obtain
\begin{equation}
	S[g,\phi,\hat{f}]=\frac{\tilde{M}_{\rm p}^2}{M_{\rm p}^2}S_{\rm EH}[g]+ S_{0}[g-\beta (\bar{M}_{\rm p}/\tilde{M}_{\rm p})^2\hat{f},\phi]+S_{2}[g,\hat{f}]\,,\label{action-spin-2-compl}
\end{equation}
where 
\begin{eqnarray}
S_{2}[g,\hat{f}]&=& -\frac{1}{4}\left(\frac{\bar{M}_{\rm p}^4}{\tilde{M}_{\rm p}^2}\beta^2 \right)\,S_{\rm FP}[g,\hat{f}] \nonumber\\[2mm]
&&- \frac{1}{4}\left(\frac{\bar{M}_{\rm p}^4}{\tilde{M}_{\rm p}^2}\beta^2 \right)\int {\rm d}^4x\sqrt{-g}\left[\left(\Lambda-\frac{R}{2}\right)\left(\hat{f}_{\mu\nu}\hat{f}^{\mu\nu}-\frac{1}{2}\hat{f}^2\right)+\Big(2\hat{f}_{\mu\rho}\hat{f}^{\rho}_\nu-\hat{f}_{\mu\nu}\hat{f} \Big)R^{\mu\nu}  \right] \nonumber\\[2mm]
&&-\frac{1}{16}\left(\frac{\bar{M}_{\rm p}^6}{\tilde{M}_{\rm p}^2}\beta^2 \right)\int {\rm d}^4x\sqrt{-g} \left[5\hat{f}_{\mu\nu}\hat{f}^{\mu\nu}\hat{f}-4\hat{f}_{\mu\nu}\hat{f}^{\mu\rho}\hat{f}_{\rho}^\nu-\hat{f}^3\right]\nonumber\\[2mm]
&&+\frac{1}{3}\left(\frac{\bar{M}_{\rm p}}{\tilde{M}_{\rm p}}\right)^4\left(\frac{\bar{M}_{\rm p}}{M_{\rm p}}\right)^2 \beta^3\int {\rm d}^4x\,{\rm d}^4y\,{\rm d}^4z\,\frac{\delta^{(3)}S_{\rm EH}}{\delta g_{\mu\nu}(x)g_{\rho\sigma}(y)g_{\alpha\beta}(z)} \hat{f}_{\mu\nu}(x)\hat{f}_{\rho\sigma}(y)\hat{f}_{\alpha\beta}(z)   \nonumber\\[2mm]
&&+\,\mathcal{O}(\hat{f}^4)
\,.\label{action-spin-2}
\end{eqnarray}
The first term is the covariant Fierz-Pauli action for the massive spin-$2$ field $\hat{f}_{\mu\nu}$ given by
\begin{eqnarray}
S_{\rm FP}[g,\hat{f}]\!\!\!&=& \!\!\!\int {\rm d}^4x\sqrt{-g}\left[-\frac{1}{2}\nabla_\rho \hat{f}_{\mu\nu}\nabla^{\rho}\hat{f}^{\mu\nu}+\nabla_\rho \hat{f}_{\mu\nu}\nabla^\mu \hat{f}^{\rho\nu} +\frac{1}{2}\nabla_{\rho}\hat{f}\nabla^\rho\hat{f} -\nabla_\mu \hat{f} \nabla_\nu\hat{f}^{\mu\nu}\right. \nonumber \\[2mm]
&&\qquad\qquad\qquad \left.-\frac{m_2^2}{2} \left(\hat{f}^{\mu\nu}\hat{f}_{\mu\nu}-\hat{f}^2\right)\right]\,,
\label{FP-action-varphi}
\end{eqnarray}
where we have defined the mass of the spin-$2$ field,
\begin{eqnarray}
m_2^2\equiv \frac{\tilde{M}_{\rm p}^2}{\beta}=\frac{\bar{M}_{\rm p}^2}{\beta}+\frac{2}{3}\Lambda=\frac{M_{\rm p}^2}{\beta}+\frac{2}{3}\Big(2\frac{\alpha}{\beta}+1\Big)\Lambda\,,
\label{spin-2-mass}
\end{eqnarray}
which, unlike the spin-$0$ mass $m_0$, depends on $\Lambda$. It is worth mentioning that in spacetime dimensions $D\neq 4$ also $m_0$ depends on $\Lambda$, so  $D=4$ is rather special in this respect~\cite{Tekin:2016vli}.
The contributions in the second line of eq.~\eqref{action-spin-2} are non-minimal interaction couplings
between $g_{\mu\nu}$ and $\hat{f}_{\mu\nu}$. Whereas, the third and fourth lines take into account cubic self-interaction, in particular the latter gives the same two-derivative cubic vertex of Einstein's GR up to a numerical coefficient. The last term $\mathcal{O}(\hat{f}^4)$ stands for higher-order interactions.

\paragraph{Fixing signs.} Similarly to the spin-$0$ case, while diagonalizing the kinetic term we also need to shift the coefficient in front of the Einstein-Hilbert action when the cosmological constant is non-zero, i.e., we have $\bar{M}_{\rm p}^2\rightarrow \tilde{M}_{\rm p}^2.$ The sign of $\tilde{M}_{\rm p}^2$ determines the overall signs of the Einstein-Hilbert and Fierz-Pauli actions in eqs.~\eqref{action-spin-2-compl} and~\eqref{action-spin-2}. Since we have to canonically normalize the fields, it is important to  fix these signs first.

To avoid a ghost-like massless graviton from $S_{\rm EH}$ we must require the condition
\begin{eqnarray}
\tilde{M}_{\rm p}^2=\bar{M}_{\rm p}^2+\frac{2}{3}\Lambda\beta>0\,. 	\label{gamma-tilde>0}
\end{eqnarray}
This also implies that $S_{\rm FP}$ is multiplied by a negative overall coefficient, which means that the additional spin-$2$ field $\hat{f}_{\mu\nu}$ is a ghost. Note that there is no way to make the coefficients of $S_{\rm EH}$ and $S_{\rm FP}$ simultaneously positive.

To prevent the massive spin-$2$ from being tachyonic we must require that $m_2^2\geq 0$ which, combined with eq.~\eqref{gamma-tilde>0}, gives
\begin{eqnarray}
\beta >0\,.	\label{beta>0}
\end{eqnarray}
Moreover,  using $\Lambda \geq 0$ and $\bar{M}_{\rm p}^2>0$ we obtain the following constraint:
\begin{eqnarray}
	m_2^2\geq \frac{2}{3}\Lambda\,.	\label{higuchi-like}
\end{eqnarray}
Therefore, a positive cosmological constant puts a lower bound on the allowed mass of the additional spin-$2$ field. It is worth mentioning that a similar inequality appears in the context of massive gravity~\cite{Hinterbichler:2011tt,deRham:2014zqa} where it is known as Higuchi bound~\cite{Higuchi:1986py} and prevents the helicity-zero mode of the massive graviton from being ghost-like around the de Sitter background. Despite having a similar expression, in our case the bound~\eqref{higuchi-like} has a more general meaning.

\paragraph{Canonical normalization.} We can now canonically normalize the field as
\begin{eqnarray}
f_{\mu\nu}= \frac{\beta}{2}\frac{\bar{M}_{\rm p}^2}{\tilde{M}_{\rm p}}\hat{f}_{\mu\nu}\,,	\label{field-fmunu}
\end{eqnarray}
thus the actions~\eqref{action-spin-2-compl} and~\eqref{action-spin-2} become
\begin{equation}
	S[g,\phi,f]=\frac{\tilde{M}_{\rm p}^2}{M_{\rm p}^2}S_{\rm EH}[g]+ S_{0}[g-2f/\tilde{M}_{\rm p},\phi]+S_{2}[g,f]\,,\label{action-spin-2-compl-2}
\end{equation}
and
\begin{eqnarray}
	S_{2}[g,f]&=&- S_{\rm FP}[g,f] - \int {\rm d}^4x\sqrt{-g}\left[\left(\Lambda-\frac{R}{2}\right)\left(f_{\mu\nu}f^{\mu\nu}-\frac{1}{2}f^2\right)+\Big(2f_{\mu\rho}f^{\rho}_\nu-f_{\mu\nu}f \Big)R^{\mu\nu}  \right]\nonumber\\[2mm]
	&&-\frac{1}{2}\frac{m_2^2}{\tilde{M}_{\rm p}} \int {\rm d}^4x\sqrt{-g} \left[5f_{\mu\nu}f^{\mu\nu}f-4f_{\mu\nu}f^{\mu\rho}f_{\rho}^\nu-f^3\right]\nonumber\\[2mm]
	&&+\frac{8}{3}\frac{1}{M_{\rm p}^2}\frac{1}{\tilde{M}_{\rm p}} \int {\rm d}^4x\,{\rm d}^4y\,{\rm d}^4z\,\frac{\delta^{(3)}S_{\rm EH}}{\delta g_{\mu\nu}(x)g_{\rho\sigma}(y)g_{\alpha\beta}(z)} f_{\mu\nu}(x)f_{\rho\sigma}(y)f_{\alpha\beta}(z)   \nonumber\\[2mm]
	&&+\,\mathcal{O}(f^4)
	\,.\label{action-spin-2-2}
\end{eqnarray}
To identify the correct dependences of the interaction couplings on $\alpha,$ $\beta$ and $\Lambda,$ for completeness we can write the explicit form of the higher-order interaction terms which are of two types:
\begin{eqnarray}
	\!\!\!\!\!\!\!\!\!\!\!\!\!\!&&\!\!\!\!\!\!\frac{(-1)^{n-2}2^{n-3}}{(n-2)!}\frac{1}{\beta}\left(\frac{1}{\tilde{M}_{\rm p}}\right)^{\!
		n-4}\!\!\int {\rm d}^4x_1\dots {\rm d}^4x_{n-2}\times\nonumber\\[2mm]
	\!\!\!\!\!\!\!\!\!\!\!\!\!\!&&\!\!\!\!\!\!\times \frac{\delta^{(n-2)}}{\delta g_{\mu_1\nu_1}(x_1)\cdots \delta g_{\mu_{n-2}\nu_{n-2}}(x_{n-2})}\!\left[\int \!{\rm d}^4x\sqrt{-g}\left(f^{\mu\nu}f_{\mu\nu}-f^2\right)\right]\!f_{\mu_1\nu_1}(x_1)\cdots f_{\mu_{n-2}\nu_{n-2}}(x_{n-2})\,,\,\,\,
	\label{int-mass-term-f}
\end{eqnarray}
and
\begin{eqnarray}
	\frac{(n-1)(-1)^{n-1}2^n}{n!\,M_{\rm p}^2}\left(\frac{1}{\tilde{M}_{\rm p}}\right)^{\!n-2}\!\!\!\int \!{\rm d}^4x_1\dots {\rm d}^4x_{n} \frac{\delta^{(n)}S_{\rm EH}}{\delta g_{\mu_1\nu_1}(x_1)\cdots \delta g_{\mu_{n}\nu_{n}}(x_{n})} f_{\mu_1\nu_1}(x_1)\cdots f_{\mu_{n}\nu_{n}}(x_{n})\,,
	\label{int-SEH-term-f}
\end{eqnarray}
where $n\geq 3$ counts the number of $f_{\mu\nu}$ in the vertex. The interactions~\eqref{int-mass-term-f} are generated by the expansion of the mass term in eq.~\eqref{action-aux-spin-2}, while those in~\eqref{int-SEH-term-f} come from the variations of the Einstein-Hilbert action.

\paragraph{Remark.} It is worth mentioning that the field redefinitions~\eqref{conformal-transf} and~\eqref{metric-shift} do not affect the functional integral because they are ultra-local, i.e., they only depend on the fields. This can be easily understood by working in dimensional regularization: the Jacobian of these transformations is equal to one because it is given by the exponential of a term proportional to $\delta^{(4)}(0)$ which vanishes in dimensional regularization (the same reasoning is also valid for local transformations which depend polynomially on the derivatives). This argument supports the fact that the renormalizability of the theory remains valid even if the equivalent form~\eqref{action-spin-2-compl-2} of the gravitational action~\eqref{action-stelle} makes it less manifest. It would be interesting to show explicitly that the UV divergences can be absorbed through additional field renormalizations which are perturbatively local by using general field-covariant methods~\cite{Anselmi:2012aq} as pointed out in~\cite{Anselmi:2018tmf}.

\subsection{Interaction couplings}\label{sec:int-coup}

When the cosmological constant is non-zero the interaction couplings acquire additional dependences on $\alpha,$ $\beta$ and $\Lambda.$ In particular, the massive spin-$2$ ghost mediates interactions whose couplings are proportional to $m_2^2=\tilde{M}_{\rm p}^2/\beta$ and  powers of $1/\tilde{M}_{\rm p}.$
Whereas,  the interactions mediated by the spin-$0$ field are proportional to $m_0^2=M^2_{\rm p}/\alpha$ and powers of $1/\bar{M}_{\rm p}.$

The shift of the Planck mass due to the presence of a cosmological constant also affects the canonical normalization and the interaction couplings of the massless graviton. Indeed, by expanding the Quadratic Gravity action in terms of
\begin{equation}
	g_{\mu\nu}=\bar{g}_{\mu\nu}+\frac{2}{\tilde{M}_{\rm p}}\,h_{\mu\nu}\,, \label{metric-fluct}
\end{equation}
where $\bar{g}_{\mu\nu}$ is a background metric and $h_{\mu\nu}$ the graviton field, we find that the self-interactions of the massless graviton are proportional to powers of $1/\tilde{M}_{\rm p}$.

We can also ask how the interaction couplings between matter and gravitational degrees of freedom look like. If $\Psi$  is a generic matter field coupled to gravity,  then the total action becomes
\begin{eqnarray}
S^\prime[g,\phi,f;\Psi] &=& \frac{\tilde{M}_{\rm p}^2}{M_{\rm p}^2}S_{\rm EH}[g]+S_0[g-2f/\tilde{M}_{\rm p},\phi]+S_2[g,f]\nonumber\\[2mm]
&&+S_{\rm m}\Big[e^{\sqrt{\frac{2}{3}}\phi/\bar{M}_{\rm p}}(g-2f/\tilde{M}_{\rm p}),\Psi\Big]\,, \label{action-grav+matter}
\end{eqnarray}
where $S_{\rm m}$ is the matter action for $\Psi$. To identify all the couplings entering in the gravity-matter interaction vertices we need to expand $g_{\mu\nu},$ $\phi,$ and $f_{\mu\nu}$ around their backgrounds which can be respectively chosen to be $\bar{g}_{\mu\nu},$ $\bar{\phi}=0,$ and $\bar{f}_{\mu\nu}=0.$ By doing so, we obtain the schematic expression
\begin{eqnarray}
S_{\rm m}[(1+\sqrt{2/3}\phi/\bar{M}_{\rm p}+\cdots)(\bar{g}+2(h-f)/\tilde{M}_{\rm p}),\Psi]\,, \label{action-matter}
\end{eqnarray}
from which it is clear that all the couplings are proportional to powers of $1/\bar{M}_{\rm p}$ and/or $1/\tilde{M}_{\rm p}.$ 
Also in this case, the interesting observation to make is that when $\Lambda= 0$ the couplings are independent of $\alpha$ and $\beta,$ whereas when $\Lambda\neq 0$ the interaction vertices acquire additional dependences on the cosmological constant and the quadratic-curvature coefficients.

\subsection{Remarks on the degrees of freedom}
 
The total number of degrees of freedom described by the action~\eqref{action-stelle} is equal to eight: a massless graviton (with $\pm 2$ helicities), a massive spin-$0$, and a massive spin-$2$ (with $\pm 2, \pm 1, 0$ helicities). By looking at the field equations of $f_{\mu\nu}$ and $g_{\mu\nu}$ we can indeed verify that $f_{\mu\nu}$ describes a massive spin-$2$ field with five helicities.
The field equations read
\begin{eqnarray}
	\frac{\delta S}{\delta f^{\mu\nu}}=0\qquad \Leftrightarrow\qquad \tilde{M}_{\rm p}\left(G_{\mu\nu}+\Lambda g_{\mu\nu}\right)=m_2^2\left(f_{\mu\nu}-g_{\mu\nu}f\right)\,,\label{EOM-f}
\end{eqnarray}
and
\begin{eqnarray}
	\!\!\!\!\!\!\!\!\!\!\!\!\!\!\frac{\delta S}{\delta g^{\mu\nu}}=0\,\,& \Leftrightarrow&\,\, \frac{\tilde{M}_{\rm p}^2}{2}\left(G_{\mu\nu}+\Lambda g_{\mu\nu}\right) +\frac{\tilde{M}_{\rm p}}{2}\left[\nabla_\rho\nabla_\mu f^\rho_\nu+\nabla_\rho\nabla_\nu f^\rho_\mu -\Box f_{\mu\nu}-g_{\mu\nu}\nabla_\rho\nabla_\sigma f^{\rho\sigma}\right.\nonumber \\[2mm]
	&& -\nabla_\mu\nabla_\nu f+g_{\mu\nu}\Box f + g_{\mu\nu}\left(G_{\rho\sigma}+\Lambda g_{\rho\sigma}\right)f^{\rho\sigma}-2\left(G_{\mu\rho}+\Lambda g_{\mu\rho}\right)f^\rho_\nu \nonumber \\[2mm]
	&& \left.-2\left(G_{\nu\rho}+\Lambda g_{\nu\rho}\right)f^\rho_\mu + \left(2\Lambda-R\right)f_{\mu\nu}+R_{\mu\nu}f\, \right]\nonumber \\[2mm]
	&& -\frac{m_2^2}{4}\left[g_{\mu\nu}\left(f_{\rho\sigma}f^{\rho\sigma}-f^2 \right)-4\left(f_{\mu}^\rho f_{\rho\nu}- f f_{\mu\nu}\right)\right]=\frac{1}{2}T^{\phi}_{\mu\nu}\,,	
	\label{EOM-gmunu}
\end{eqnarray}
where we have defined the stress-energy tensor
\begin{eqnarray}
T^{\phi}_{\mu\nu}=-\frac{2}{\sqrt{-g}}\frac{\delta S_{0}}{\delta g^{\mu\nu}}\,.\label{stress-phi}
\end{eqnarray}
If we take the covariant divergence of~\eqref{EOM-f} we obtain the following constraint equation:
\begin{eqnarray}
	\nabla_\mu f^\mu_\nu=\nabla_\nu f\,,
	\label{constraint-f}
\end{eqnarray}
which kills four degrees of freedom, i.e., we get $10-4=6.$ Moreover, if we take the trace of~\eqref{EOM-gmunu} and use~\eqref{constraint-f} we obtain
\begin{eqnarray}
	\tilde{M}_{\rm p}\left(m_2^2-\frac{2}{3}\Lambda\right) f=-3\,T^{\phi}\,,
	\label{trace-gEOM}
\end{eqnarray}
where $T^{\phi}=g^{\mu\nu}T^{\phi}_{\mu\nu}.$ At the linear level the trace equation reads $\tilde{M}_{\rm p}\left(\frac{2}{3}\Lambda-m_2^2\right) f= 0$ as $T^{\phi}\sim \mathcal{O}(\phi^2),$ thus if $m_2^2\neq \frac{2}{3}\Lambda$ we get the additional constraint $f=0.$ This means that five of the ten components of $f_{\mu\nu}$ are killed, i.e., $10-4-1=5,$ implying that $f_{\mu\nu}$ is a massive spin-$2$ field.

However, it is extremely important to emphasize that this type of counting of the degrees of freedom is valid as long as the condition $m_2^2\neq\frac{2}{3}\Lambda$ holds true. If, instead, $m_2^2= \frac{2}{3}\Lambda$ the linearized trace equation vanishes \textit{identically} suggesting that some degrees of freedom are redundant. This happens in both the massless $m_2^2=0$ (when $\Lambda=0$) and partially massless $m_2^2=\frac{2}{3}\Lambda$ (when $\Lambda>0$) scenarios where enhanced gauge symmetries arise. In such situations further work is needed to keep track of all degrees of freedom, and the detailed structure of the particle spectrum can be shown to change, although the total number remains eight.

Interestingly, these two special cases arise in the limit $\beta \rightarrow \infty$ of the theory~\eqref{action-stelle} due to the peculiar dependence on $\beta$ and $\Lambda$ of the spin-$2$ mass $m_2$. In the next section we will explicitly find the gravitational theory resulting in the limit $\beta \rightarrow \infty$ in both cases $\Lambda=0$ and $\Lambda>0$ taking carefully into account the detailed structure of the particle spectrum and gauge symmetries.

\section{Limit $\beta\rightarrow \infty$ of Quadratic Gravity}\label{sec:decouplings}

We are interested in the limit $\beta\rightarrow \infty$ of the gravitational action in eq.~\eqref{action-stelle} with all other parameters and the canonically normalized fields kept fixed. As already briefly mentioned, this limit corresponds to special values of the spin-$2$ ghost mass $m_2:$
\begin{equation}
\beta\rightarrow \infty \quad \Leftrightarrow \quad \left\lbrace \begin{array}{cc} \!\!\!\!\!m_2^2\rightarrow 0\,,& \text{if}\,\,\, \Lambda =0\,;\\[2mm]
\displaystyle m_2^2\rightarrow \frac{2}{3}\Lambda\,,& \text{if}\,\,\,  \Lambda> 0\,. \label{mass-limits}
\end{array}\right.
\end{equation}
In the first case we have to study the \textit{massless limit} of a massive spin-$2$ field; whereas, when the cosmological constant is non-zero and positive we have to consider a \textit{partially massless limit}~\cite{Deser:1983mm,Deser:1983tm,Deser:2001us,DeRham:2018axr}. These types of limits have been intensively studied in the context of ghost-free theories of massive gravity~\cite{Hinterbichler:2011tt,deRham:2014zqa,DeRham:2018axr}. In the massless limit the massive spin-$2$ splits into a massless spin-$2,$ a massless spin-$1$ and a massless spin-$0$ such that the total number of degrees of freedom is still five ($2+2+1=5$). In the partially massless limit the splitting is different: the massive spin-$2$ splits into a partially massless spin-$2$ graviton (which only has the $\pm2$ and $\pm 1$ helicities due to an additional scalar gauge symmetry)~\cite{Deser:1983mm,Deser:1983tm,Deser:2001us} and a massive spin-$0,$ thus also in this case the total number of degrees of freedom is still five ($4+1=5$). 

Therefore, investigating the limit $\beta\rightarrow \infty$ of Quadratic Gravity is equivalent to analysing the massless or partially massless limits depending on whether the cosmological constant is zero or not. This means that we cannot simply work with the action~\eqref{action-spin-2-compl-2} because in the two limits~\eqref{mass-limits} the details of the particle spectrum change. A useful method to account for all the helicities  in a consistent way  when taking the two limits is the St\"uckelberg formalism. 

An aspect to highlight is that in standard ghost-free theories of massive gravity the mass limits~\eqref{mass-limits} always hit a strong coupling regime because of the helicity-zero interactions whose couplings scale as powers of $1/m_2$ when $\Lambda=0$ or as $1/(m_2^2-\frac{2}{3}\Lambda)$ when $\Lambda\neq 0$~\cite{Hinterbichler:2011tt,deRham:2014zqa}. However, here we will show that the property of renormalizability (in four dimensions) and the ghost-like nature of the massive spin-$2$ field together guarantee that the limit $\beta\rightarrow \infty$ is regular, i.e., no strong coupling arises. We will consider the scenarios $\Lambda=0$ and $\Lambda> 0$ separately and  show that only in $D=4$ the strong coupling can be avoided.

\subsection{Case $\Lambda=0$: $m^2_2\rightarrow 0$ (massless limit)}\label{sec:beta-inf-lambda=0}

Let us generalize the action~\eqref{action-weyl->Ricci} to $D$ spacetime dimensions and set $\Lambda=0$ in this subsection:
\begin{equation}
	S[g,\phi]=S_{\rm EH}[g]\big|_{\Lambda=0}+ S_{0}[g,\phi]-\frac{\beta}{2}\int {\rm d}^D x\sqrt{-g}\left(R_{\mu\nu}R^{\mu\nu}-\frac{D}{4(D-1)}R^2\right)\,,\label{action-weyl->Ricci-D}
\end{equation}
where the factor $\frac{D}{4(D-1)}$ makes sure that the terms multiplied by $\beta$ only contribute to the spin-$2$ component of the gravitational multiplet. Introducing the canonically normalized  field $f_{\mu\nu}$ we can rewrite the action as
\begin{equation}
	S[g,\phi,f]=S_{\rm EH}[g]\big|_{\Lambda=0}+ S_{0}[g,\phi]-\int {\rm d}^D x\sqrt{-g}\left[M_{\rm p}^{\frac{D-2}{2}}G_{\mu\nu}f^{\mu\nu}-\frac{m_2^2}{2}(f_{\mu\nu}f^{\mu\nu}-f^2) \right]\,,\label{action-spin-2-D}
\end{equation}
where $f_{\mu\nu}=\frac{M_{\rm p}^{(D-2)/2}}{m_2^2}[R_{\mu\nu}-\frac{1}{2(D-1)}g_{\mu\nu}R].$

\paragraph{St\"uckelberg formalism.} Since we are interested in the massless limit it is convenient to use the St\"uckelberg formalism~\cite{Hinterbichler:2011tt,deRham:2014zqa} to keep track of all five helicities of the massive spin-$2$ field. The idea is to write an equivalent action in which the spin-$2$ field $f_{\mu\nu}$ manifestly exhibits a gauge symmetry. This can be done by introducing new fields through the replacement
\begin{equation}
f_{\mu\nu}\rightarrow f_{\mu\nu}+\nabla_\mu V_\nu+\nabla_\nu V_\mu\,,\qquad V_{\mu}=\hat{A}_{\mu}+\nabla_\mu \hat{\chi}\,,\label{f-stuck}
\end{equation}
such that the following gauge symmetry is satisfied:
\begin{equation}
\delta_\xi f_{\mu\nu}=\nabla_\mu \xi_\nu +\nabla_\nu \xi_\mu\,,\qquad \delta_\xi \hat{A}_\mu=-\xi_\mu+\nabla_\mu \xi\,,\qquad \delta_\xi \hat{\chi}=-\xi\,, \label{gauge transf.}
\end{equation}
$\xi_\mu$ and $\xi$ being an arbitrary gauge vector and an arbitrary gauge scalar function, respectively. The new  action reads
\begin{eqnarray}
	S[g,\phi,f,V]&=&S_{\rm EH}[g]\big|_{\Lambda=0}+ S_{0}[g,\phi]-\int {\rm d}^D x\sqrt{-g}\left[M_{\rm p}^{\frac{D-2}{2}}G_{\mu\nu}f^{\mu\nu}-\frac{m_2^2}{2}(f_{\mu\nu}f^{\mu\nu}-f^2)\right.\nonumber \\[2mm]
&&\qquad\left.	-2m_2^2(f^{\mu\nu}\nabla_\mu V_\nu-f\nabla_\mu V^\mu)-\frac{m_2^2}{2}\hat{F}_{\mu\nu}\hat{F}^{\mu\nu}+2m_2^2R_{\mu\nu}V^{\mu}V^\nu \right]\,,
	\label{action-spin-2-D-stuck}
\end{eqnarray}
where $\hat{F}_{\mu\nu}=\nabla_\mu V_\nu-\nabla_\nu V_\mu=\nabla_\mu \hat{A}_\nu-\nabla_\nu \hat{A}_\mu.$ The action~\eqref{action-spin-2-D-stuck} with $m_2\neq 0$ is equivalent to~\eqref{action-spin-2-D} because we can always choose a gauge in which the fields $\hat{A}_\mu$ and $\hat{\chi}$ are zero.

\paragraph{Regularity of the limit.} Note that after canonical normalization,  $\hat{A}_\mu\sim \frac{1}{m_2}A_\mu$ and $\hat{\chi} \sim \frac{1}{m_2^2}\chi,$ the term  $m_2^2R_{\mu\nu}V^{\mu}V^\nu$ gives the contributions $\frac{1}{m_2}R_{\mu\nu}A^\mu\nabla^\nu \chi$ and $\frac{1}{m_2^2}R_{\mu\nu}\nabla^\mu \chi\nabla^\nu \chi$ which diverge in the limit $m_2\rightarrow 0$. In addition, the diagonalization of the kinetic operator for $f_{\mu\nu}$ and $\hat{\chi}$ could introduce strongly coupled interaction terms for the helicity-zero.
However,  we now show that in four spacetime dimensions the property of renormalizability and the ghost-like nature of the massive spin-$2$ allow for cancellations that make the massless limit regular; to our knowledge this feature was first noticed in~\cite{Hinterbichler:2015soa}.

Let us perform the  field transformation
\begin{equation}
f_{\mu\nu}\rightarrow f_{\mu\nu}+a\,V_\mu V_\nu+b\,g_{\mu\nu}V^\rho V_\rho\,, \label{f-transf.}
\end{equation}
and check whether there exist values of the two constants $a$ and $b$ such that the (apparent) divergent terms cancel out. The action becomes
\begin{eqnarray}
	S[g,\phi,f,V]\!\!&\!\!=\!&\!\!S_{\rm EH}[g]\big|_{\Lambda=0}+ S_{0}[g,\phi]-\!\int\! {\rm d}^D x\sqrt{-g}\left\lbrace M_{\rm p}^{\frac{D-2}{2}}G_{\mu\nu}f^{\mu\nu}+\left[a M_{\rm p}^{\frac{D-2}{2}}+2m_2^2\right]R_{\mu\nu}V^\mu V^\nu\right.\nonumber \\[2mm]
\!\!\!	&&\!\!\!+M_{\rm p}^{\frac{D-2}{2}}\left[b\left(1-\frac{D}{2}\right)-\frac{a}{2}\right]R V_\rho V^\rho-\frac{m_2^2}{2}\left(f_{\mu\nu}f^{\mu\nu}-f^2\right)-a\,m_2^2f^{\mu\nu} V_\mu V_\nu	\nonumber\\[2mm]
\!\!\!	&&\!\!\!-\frac{m_2^2}{2}[2b(1-D)-2a]f V_\rho V^\rho -\frac{m_2^2}{2}b(bD+2a)(1-D)(V_\rho V^\rho)^2-\frac{m_2^2}{2}\hat{F}_{\mu\nu}\hat{F}^{\mu\nu}\nonumber\\[2mm]
\!\!\!	&&\!\!\!-2m_2^2(f^{\mu\nu}\nabla_\mu V_\nu-f\nabla_\rho V^\rho)-2m_2^2[2b(1-D)-3a]V_\rho V^\rho\nabla_\mu V^\mu
	 \Big\rbrace \,,
	\label{action-spin-2-D-transf}
\end{eqnarray}
where the terms that can blow up in the massless limit after canonical normalization are now given by $m_2^2(V_{\rho}V^\rho)^2,$ $m_2^2 V^\rho V_\rho\nabla_\mu V^\mu,$ $R_{\mu\nu} V^\mu V^\nu$ and $R V_\rho V^\rho.$ Therefore, we must fix the values of $a$ and $b$ such that these terms disappear from the action, i.e., we must impose
\begin{eqnarray}
\left\lbrace \begin{array}{rl} 
\displaystyle a M_{\rm p}^{\frac{D-2}{2}}+2m_2^2=0\,,	&\\[2mm]
\displaystyle b(2-D)-a=0\,,&\\[2mm]
\displaystyle b(bD+2a)(1-D)=0\,,&\\[2mm]
\displaystyle 2b(1-D)-3a=0\,.&
\end{array}  \right.
\label{set-conditions}
\end{eqnarray}
These equations can be simultaneously solved \textit{only} in $D=4$ spacetime dimensions. Note that the theory of Quadratic Gravity in eq.~\eqref{action-stelle} is renormalizable only in four dimensions~\cite{Anselmi:2019xac}. This suggests that the property of perturbative renormalizability guarantees that the massless limit does not hit a strong coupling.  However, it should be emphasized that renormalizability is not the only key feature behind these cancellations, but the ghost-like nature of the spin-$2$ field also plays a crucial role\footnote{It is worth mentioning that a similar feature may also characterize some massive Yang-Mills models that are renormalizable and contain ghost-like degrees of freedom. An example is given by the Curci-Ferrari model~\cite{Curci:1976bt,Curci:1976kh} in which the massless limit is regular.}. Indeed, the interaction term $G_{\mu\nu}f^{\mu\nu}$ is important for our derivation and its structure requires that one of the two spin-$2$ fields ($h_{\mu\nu}$ or $f_{\mu\nu}$) is ghost-like depending on the overall multiplicative sign; in our case $h_{\mu\nu}$ is chosen to be a normal graviton, while $f_{\mu\nu}$ a ghost.

\paragraph{Resulting theory.} We now substitute the solution $D=4$ and $a=-2b=-2m_2^2/M_{\rm p}$ of the system~\eqref{set-conditions} into the action~\eqref{action-spin-2-D-transf} and canonically normalize the fields as
\begin{eqnarray}
A_\mu=\sqrt{2}m_2 \hat{A}_\mu\,,\qquad \chi = \sqrt{6}m_2^2 \hat{\chi}\,,
	\label{canonic-norm-A-pi}
\end{eqnarray}
thus the limit $m_2^2\rightarrow 0$ of~\eqref{action-spin-2-D-transf} reads
\begin{eqnarray}
\!\!\!\!\!\!\!\!\!\!	S[g,\phi,f,A,\chi]\!\!\!&=&\!\!\!S_{\rm EH}[g]\big|_{\Lambda=0}+ S_{0}[g,\phi]-\int {\rm d}^4 x\sqrt{-g}\left[ M_{\rm p}G_{\mu\nu}f^{\mu\nu}-\frac{1}{4}F_{\mu\nu}F^{\mu\nu}\right.\nonumber \\[2mm]
	&&\left.\!\!\!	+\frac{1}{3M_{\rm p}}f^{\mu\nu}\left(\nabla_\mu\chi\nabla_\nu\chi +\frac{1}{2}g_{\mu\nu}\nabla_\rho\chi\nabla^\rho\chi \right)  -\sqrt{\frac{2}{3}}f^{\mu\nu}\left(\nabla_\mu\nabla_\nu\chi-g_{\mu\nu}\Box\chi \right)\right] \,,
	\label{action-spin-2-V-chi}
\end{eqnarray}
where  $F_{\mu\nu}=\nabla_\mu A_\nu-\nabla_\nu A_\mu,$  $\Box=g^{\mu\nu}\nabla_\mu\nabla_\nu$ is the d'Alembertian operator, and with an abuse of notation we called the limit of the action, i.e.,  $\lim\limits_{\beta\rightarrow \infty} S$, with the same name $S$.   

The gauge transformations~\eqref{gauge transf.} are now given by
\begin{equation}
	\delta_\xi f_{\mu\nu}=\nabla_\mu \xi_\nu +\nabla_\nu \xi_\mu+\sqrt{\frac{2}{3}}\frac{1}{M_{\rm p}}\left[\xi_\mu\nabla_\nu\chi+\xi_\nu\nabla_\mu\chi-g_{\mu\nu}\xi_\rho\nabla^\rho\chi\right]\,,\quad\,\, \delta_\xi A_\mu=\nabla_\mu \xi\,,\quad\,\, \delta_\xi \chi=0\,; \label{gauge transf-2}
\end{equation}
the helicity-zero becomes gauge invariant in the massless limit. 

The next step is to diagonalize the kinetic operator for $f_{\mu\nu}$ and $\chi.$ This can be done by first performing the Weyl transformation
\begin{equation}
g_{\mu\nu}\rightarrow e^{-\sqrt{\frac{2}{3}}\chi/M_{\rm p}}g_{\mu\nu}\,,
\end{equation}
and then shifting $f_{\mu\nu}$ as
\begin{equation}
f_{\mu\nu}\rightarrow f_{\mu\nu}+\frac{M_{\rm p}}{2}\left(1- e^{-\sqrt{\frac{2}{3}}\chi/M_{\rm p}} \right)g_{\mu\nu}\,,
\end{equation}
thus the action~\eqref{action-spin-2-V-chi} becomes
\begin{eqnarray}
	S[g,f,A,\phi,\chi] =S_{\rm EH}[g]\big|_{\Lambda=0}-M_{\rm p}\int {\rm d}^4 x\sqrt{-g}G_{\mu\nu}f^{\mu\nu}+S_1[g,A]+S_{00}[g,\phi,\chi]\,,
	\label{action-spin-2-V-chi-3}
\end{eqnarray}
where we have used $G_{\mu\nu}(g)+\frac{1}{3M^2_{\rm p}}(\nabla_\mu\chi\nabla_\nu\chi+\frac{1}{2}g_{\mu\nu}\nabla_\rho\chi\nabla^\rho\chi)-\sqrt{2/3}\frac{1}{M_{\rm p}}(\nabla_\mu\nabla_\nu\chi-g_{\mu\nu}\Box\chi)=G_{\mu\nu}\big(e^{\sqrt{2/3}\chi/M_{\rm p}}g\big),$ and defined
\begin{eqnarray}
	S_1[g,A]=\frac{1}{4}\int {\rm d}^4x\sqrt{-g} F_{\mu\nu}F^{\mu\nu}
	\label{action-spin-1}
\end{eqnarray}
and
\begin{eqnarray}
	S_{00}[g,\phi,\chi]\!\!&\!\!=\!\!&\!\! \int {\rm d}^4x\sqrt{-g}\left[e^{-\sqrt{\frac{2}{3}}\chi/M_{\rm p}}\left(-\frac{1}{2}\nabla_\mu\phi\nabla^\mu\phi+\frac{1}{2}\nabla_\mu\chi\nabla^\mu\chi\right)\right.\nonumber \\[2mm]
	&&\qquad\qquad\qquad\left.-\frac{m_0^2}{2}\frac{3M_{\rm p}^2}{2}e^{-2\sqrt{\frac{2}{3}}\chi/M_{\rm p}}\left(1-e^{\sqrt{\frac{2}{3}}\phi/M_{\rm p}}\right)^2 \right]\,.
	\label{S00-lambda=0}
\end{eqnarray}

The gauge transformations~\eqref{gauge transf-2} reduce to
\begin{equation}
	\delta_\xi f_{\mu\nu}=\nabla_\mu \xi_\nu +\nabla_\nu \xi_\mu\,,\qquad \delta_\xi A_\mu=\nabla_\mu \xi\,,\qquad \delta_\xi \chi=0\,. \label{gauge transf-3}
\end{equation}

Therefore, in the limit $m_2\rightarrow 0$ (i.e., $\beta\rightarrow \infty$ with $\Lambda=0$) the massive spin-$2$ ghost splits into a massless spin-$2$ field $f_{\mu\nu}$ with helicities $\pm 2,$ a massless spin-$1$ field $A_\mu$ with helicities $\pm 1,$ and a gauge invariant scalar field $\chi:$ in total we still have $2+2+1=5$ ghost-like degrees of freedom. Our result is in agreement with~\cite{Hinterbichler:2015soa}\footnote{In the last part of Ref.~\cite{Hinterbichler:2015soa} it is claimed that in Quadratic Gravity every amplitude vanishes in the high-energy limit. Unfortunately this statement is  incorrect as it was also noted later by one of the authors~\cite{Bonifacio:2018aon}. In fact, it is well-known that despite the property of renormalizability in Quadratic Gravity tree-level scattering amplitudes grow with the energy squared~\cite{Dona:2015tra,Holdom:2021hlo,Abe:2022spe}. Nevertheless, we agree with~\cite{Hinterbichler:2015soa} regarding the regularity of the massless limit.} where a similar analysis was performed in the case $\Lambda=0$.

We can diagonalize the kinetic operator of the metric $g_{\mu\nu}$ and the spin-$2$ ghost $f_{\mu\nu}$ by making the transformation $g_{\mu\nu}\rightarrow g_{\mu\nu}-2f_{\mu\nu}/M_{\rm p}.$ Then, expanding in $f_{\mu\nu}$ we obtain
\begin{eqnarray}
	S[g,\phi,f,A,\chi] &=& S_{\rm EH}[g]\big|_{\Lambda=0}+S_{2}[g,f]\big|_{m_2=0,\,\Lambda=0}+S_1[g-2f/M_{\rm p},A]\nonumber\\[2mm]
	&&+S_{00}[g-2f/M_{\rm p},\phi,\chi]\,,
	\label{action-spin-2-V-chi-4}
\end{eqnarray}
where $S_{2}[g,f]\big|_{m_2=0,\,\Lambda=0}$ is equal to the expression in eq.~\eqref{action-spin-2-2} with $m_2=0$ and $\Lambda=0.$ From eq.~\eqref{action-spin-2-V-chi-4} we can see that all the interaction couplings scale as powers of $1/M_{\rm p}$ and no degree of freedom decouples.

Although the structure of the particle spectrum -- seven massless ($\pm 2,\pm 2, \pm 1, 0$) and one massive spin-$0$ -- is similar to that of conformal gravity plus two additional scalar fields, the action~\eqref{action-spin-2-V-chi-3} (or~\eqref{action-spin-2-V-chi-4}) represents a different theory because the Planck mass $M_{\rm p}$ does not disappear in the limit and  breaks the scale invariance of $S_{\rm W}$. Another way to understand this fact is to look at eq.~\eqref{action-spin-2-D}: in the limit $m^2_2\rightarrow 0$ it becomes impossible to recast the action as $S_{\rm W}$ $+$ $S_{00}$ by integrating out $f_{\mu\nu}.$ 

It is also worth mentioning that thanks to the ghost-like nature of the additional spin-$2$ field it becomes possible to consistently couple a massless spin-$2$ field to the metric $g_{\mu\nu}$ and evade well-known no-go theorems~\cite{Ogievetsky:1965zcd,Wald:1986bj,Cutler:1986dv,Boulanger:2000rq} as already noted in~\cite{Hindawi:1995an}.

\subsection{Case $\Lambda > 0$: $m^2_2\rightarrow \frac{2}{3}\Lambda$ (partially massless limit)}\label{sec:beta-inf-lambdaneq0}

We now analyse the limit $\beta\rightarrow \infty$ when $\Lambda >0$ and show that new interesting features appear as compared to the previous scenario. Also in this case, only in $D=4$ spacetime dimensions the limit does not hit a strong coupling. In what follows, we directly work in four spacetime dimensions and show that the limit is regular.

The starting point is the gravitational action~\eqref{action-stelle} with $\Lambda\neq 0$ written in terms of the canonically normalized fields $\phi$ and $f_{\mu\nu},$
\begin{equation}
	S[g,\phi,f]=\frac{\tilde{M}_{\rm p}^2}{M_{\rm p}^2}S_{\rm EH}[g]+ S_{0}[g,\phi]-\int {\rm d}^4 x\sqrt{-g}\left[\tilde{M}_{\rm p}(G_{\mu\nu}+\Lambda g_{\mu\nu})f^{\mu\nu}-\frac{m_2^2}{2}(f_{\mu\nu}f^{\mu\nu}-f^2) \right]\,,\label{action-spin-2-lambda}
\end{equation}
where now $f_{\mu\nu}=\frac{\tilde{M}_{\rm p}}{m_2^2}[R_{\mu\nu}-\frac{1}{6}g_{\mu\nu}R-\frac{1}{3}\Lambda g_{\mu\nu}].$

\paragraph{St\"uckelberg formalism.} To consistently account for all  five helicities in the limit $m_2^2\rightarrow \frac{2}{3}\Lambda$, it is again convenient to use the St\"uckelberg trick. Now the idea is to write an equivalent action in which $f_{\mu\nu}$ manifestly exhibits the so-called partially massless symmetry~\cite{Deser:1983mm,Deser:1983tm,Deser:2001us}. To achieve this goal, we also need to restore a Weyl symmetry for the metric tensor. 

We can start by making the replacements
\begin{equation}
g_{\mu\nu}\rightarrow e^{-m^2_2\,\hat{\chi}/\tilde{M}_{\rm p}} g_{\mu\nu}\,,\qquad f_{\mu\nu}\rightarrow f_{\mu\nu}+\nabla_\mu\nabla_\nu \hat{\chi}+\frac{m_2^2}{\tilde{M}_{\rm p}}\left(\nabla_\mu\hat{\chi} \nabla_\nu\hat{\chi} - \frac{1}{2}g_{\mu\nu} \nabla_\rho \hat{\chi}\nabla^\rho \hat{\chi} \right)\,,\label{decomp-f-lambda}
\end{equation}
such that the new action is invariant under the following gauge transformations:
\begin{eqnarray}
\delta_\zeta f_{\mu\nu}\!\!&=&\!\!\nabla_\mu\nabla_\nu \zeta + \frac{m_2^2}{2\tilde{M}_{\rm p}}\left(\nabla_\mu\hat{\chi} \nabla_\nu\zeta + \nabla_\nu\hat{\chi} \nabla_\mu\zeta - g_{\mu\nu} \nabla_\rho \hat{\chi}\nabla^\rho \zeta \right)\,,\nonumber\\[2mm]
\delta_\zeta g_{\mu\nu}\!\!&=&\!\!-\frac{m_2^2}{\tilde{M}_{\rm p}}\zeta g_{\mu\nu}\,,\nonumber \\[2mm]
\delta_\zeta \hat{\chi}\!\!&=&\!\! -\zeta \,,\label{gauge-transf-lambda}
\end{eqnarray}
$\zeta(x)$ being an arbitrary scalar gauge function. The new action reads
\begin{eqnarray}
	S[g,\phi,f,\hat{\chi}]\!&=&\!\frac{\tilde{M}_{\rm p}^2}{M_{\rm p}^2}S_{\rm EH}\big[e^{-m_2^2\hat{\chi}/\tilde{M}_{\rm p}}g\big]+ S_{0}\big[e^{-m_2^2\hat{\chi}/\tilde{M}_{\rm p}}g,\phi\big]\nonumber\\[2mm]
	\!&&\!+\int {\rm d}^4 x\sqrt{-g}\left\lbrace -\tilde{M}_{\rm p}\left[G_{\mu\nu}-\frac{m_2^4}{2\tilde{M}^2_{\rm p}}\left( \nabla_\mu\hat{\chi}\nabla_\nu \hat{\chi}+\frac{1}{2}g_{\mu\nu}\nabla_\rho\hat{\chi}\nabla^\rho\hat{\chi}  \right)  \right]f^{\mu\nu} \right.\nonumber\\[2mm]
	\!&&\!\left.-\tilde{M}_{\rm p}\Lambda e^{-m_2^2\hat{\chi}/\tilde{M}_{\rm p}}f+\frac{m_2^2}{2}(f_{\mu\nu}f^{\mu\nu}-f^2) -\frac{m_2^2}{2}R^{\mu\nu}\nabla_\mu \hat{\chi}\nabla_\nu \hat{\chi} \right\rbrace\,,
	\label{action-spin-2-lambda-2}
\end{eqnarray}
which is indeed invariant under~\eqref{gauge-transf-lambda} and is equivalent to~\eqref{action-spin-2-lambda} when $m_2^2\neq \frac{2}{3}\Lambda$ because we can fix the gauge as $\hat{\chi}=0.$

We now perform the transformation
\begin{eqnarray}
	f_{\mu\nu}\rightarrow f_{\mu\nu}-\frac{m_2^2}{2\tilde{M}_{\rm p}}\left(\nabla_\mu\hat{\chi}\nabla_\nu\hat{\chi} -\frac{1}{2}g_{\mu\nu}\nabla_\rho \hat{\chi}\nabla^\rho \hat{\chi}\right)
	\label{transformation-f-lambda}
\end{eqnarray}
to make the limit $\beta\rightarrow\infty$ manifestly regular, and  shift
\begin{eqnarray}
	f_{\mu\nu}\rightarrow f_{\mu\nu}+\frac{\tilde{M}_{\rm p}}{2}\left(1-e^{-m_2^2\hat{\chi}/\tilde{M}_{\rm p}}\right)g_{\mu\nu}\,.
	\label{shift-transf-lambda}
\end{eqnarray}
to diagonalize $f_{\mu\nu}$ and $\hat{\chi}$ up to terms that vanish in the limit $\beta\rightarrow \infty$. Then, if we canonically normalize the field $\hat{\chi}$ as
\begin{eqnarray}
	\chi=\sqrt{\frac{3}{2}}m_2\sqrt{m_2^2-\frac{2}{3}\Lambda}\,\hat{\chi}=\sqrt{\frac{3}{2}}m_2\frac{\bar{M}_{\rm p}}{\sqrt{\beta}}\,\hat{\chi}\,,
	\label{canonic-pi-lambda}
\end{eqnarray}
the action~\eqref{action-spin-2-lambda-2} becomes
\begin{eqnarray}
	S[g,\phi,f,\chi]&=&\frac{\tilde{M}_{\rm p}^2}{M_{\rm p}^2}S_{\rm EH}[g]+ S_{0}\Big[e^{-\sqrt{\frac{2}{3}}\frac{m_2\sqrt{\beta}}{\bar{M}_{\rm p}\tilde{M}_{\rm p}}\chi}g,\phi\Big]\nonumber\\[2mm]
	&&+\int {\rm d}^4 x\sqrt{-g}\left[-\tilde{M}_{\rm p}(G_{\mu\nu}+\Lambda g_{\mu\nu})f^{\mu\nu}+\frac{m_2^2}{2}(f_{\mu\nu}f^{\mu\nu}-f^2) \right]\nonumber \\[2mm]
	&&    +\int {\rm d}^4x\sqrt{-g} \left[\frac{1}{2}e^{-\sqrt{\frac{2}{3}}\frac{m_2\sqrt{\beta}}{\bar{M}_{\rm p}\tilde{M}_{\rm p}}\chi}\nabla_\mu\chi\nabla^\mu\chi -\frac{3}{2}\frac{\bar{M}_{\rm p}^2}{\beta}\tilde{M}_{\rm p}^2\left(1-e^{-\sqrt{\frac{2}{3}}\frac{m_2\sqrt{\beta}}{\bar{M}_{\rm p}\tilde{M}_{\rm p}}\chi} \right)^2\right]   \nonumber \\[2mm]
	&& -\frac{3}{2}\tilde{M}_{\rm p}\frac{\bar{M}_{\rm p}^2}{\beta}\int{\rm d}^4x\sqrt{-g}f\left(1-e^{-\sqrt{\frac{2}{3}}\frac{m_2\sqrt{\beta}}{\bar{M}_{\rm p}\tilde{M}_{\rm p}}\chi} \right)  \,.
	\label{action-spin-2-lambda-4}
\end{eqnarray}
The gauge transformations~\eqref{gauge-transf-lambda} now read
\begin{eqnarray}
	\delta_\zeta f_{\mu\nu}=\nabla_\mu\nabla_\nu \zeta + \frac{m_2^2}{2}g_{\mu\nu}\zeta\,,\qquad
	\delta_\zeta g_{\mu\nu}=-\frac{m_2^2}{\tilde{M}_{\rm p}}\zeta g_{\mu\nu}\,,\qquad
	\delta_\zeta \chi= -\sqrt{\frac{3}{2}}\frac{m_2\bar{M}_{\rm p}}{\sqrt{\beta}}\zeta \,.\label{gauge-transf-lambda-2}
\end{eqnarray}
Therefore, thanks to the introduction of the St\"uckelberg field $\chi$ the action~\eqref{action-spin-2-lambda-4} remains invariant if we simultaneously transform $g_{\mu\nu}$ and $f_{\mu\nu}$ together with $\chi,$ namely the action manifestly exhibits Weyl and partially massless symmetries, simultaneously. It is worth emphasizing that the above discussion is entirely covariant, independent of the background metric and valid for any value of the mass $m_2$.

\paragraph{Resulting theory.} The limit $\beta\rightarrow \infty$ (i.e., the partially massless limit $m_2^2\rightarrow \frac{2}{3}\Lambda$) of eq.~\eqref{action-spin-2-lambda-4} is regular and, moreover, the $f\chi$ mixing term in the last line goes to zero. To fully take the limit we have to diagonalize $g_{\mu\nu}$ and $f_{\mu\nu}$ and expand in field fluctuations so that we can correctly identify all couplings of the interactions among all canonically normalized fields. 
However, before doing so, let us consider an intermediate step such that the limit $\beta\rightarrow \infty$ is taken at the level of the covariant action~\eqref{action-spin-2-lambda-4} in which $g_{\mu\nu}$ is not yet split in background part plus fluctuations. This will allow us to make a comparison with cosmological conformal gravity. 

Up to terms of order $\mathcal{O}(1/\sqrt{\beta})$ the limit $\beta\rightarrow \infty$ of~\eqref{action-spin-2-lambda-4} reads   
\begin{eqnarray}
	S[g,f,\phi,\chi]&=&\frac{2\Lambda \beta}{3M_{\rm p}^2}S_{\rm EH}[g]+S_{00}[g,\phi,\chi]\nonumber\\[2mm] 
	&&+\int {\rm d}^4 x\sqrt{-g}\left[-\sqrt{\frac{2\Lambda\beta}{3}}\,(G_{\mu\nu}+\Lambda g_{\mu\nu})f^{\mu\nu}+\frac{\Lambda}{3}(f_{\mu\nu}f^{\mu\nu}-f^2) \right]\,,\quad
	\label{action-spin-2-lambda-5}
\end{eqnarray}
where we have used $\tilde{M}_{\rm p}^2\sim \frac{2}{3}\Lambda\beta$ and defined
\begin{eqnarray}
\!\!\!\!\!\!\!	S_{00}[g,\phi,\chi]\!\!&\!\!=\!\!&\!\! \int {\rm d}^4x\sqrt{-g}\left[e^{-\sqrt{\frac{2}{3}}\chi/\bar{M}_{\rm p}}\left(-\frac{1}{2}\nabla_\mu\phi\nabla^\mu\phi+\frac{1}{2}\nabla_\mu\chi\nabla^\mu\chi\right)\right.\nonumber \\[2mm]
\!\!\!\!\!\!\!	&&\qquad\left.-\bar{M}_{\rm p}^2\Lambda\left(1-e^{-\sqrt{\frac{2}{3}}\chi/\bar{M}_{\rm p}}\right)^2-\frac{m_0^2}{2}\frac{3\bar{M}_{\rm p}^2}{2}e^{-2\sqrt{\frac{2}{3}}\chi/\bar{M}_{\rm p}}\left(1-e^{\sqrt{\frac{2}{3}}\phi/\bar{M}_{\rm p}}\right)^2 \right]\,.
	\label{S00-lambdaneq0}
\end{eqnarray}
The gauge transformations are now given by
\begin{eqnarray}
	\delta_\zeta f_{\mu\nu}=\nabla_\mu\nabla_\nu\zeta+\frac{\Lambda}{3}g_{\mu\nu}\zeta \,,\qquad \delta_\zeta g_{\mu\nu}=-g_{\mu\nu}\sqrt{\frac{2\Lambda}{3\beta}}\zeta\rightarrow 0\,,\qquad \delta_\zeta \chi=-\bar{M}_{\rm p}\sqrt{\frac{\Lambda}{\beta}}\zeta\rightarrow 0\,.
	\label{gauge-transf-lambda2}
\end{eqnarray}
Note that to show the gauge invariance of~\eqref{action-spin-2-lambda-5} under the transformations~\eqref{gauge-transf-lambda2} we must be careful about the limit $\beta\rightarrow \infty$ of $\delta_\zeta g_{\mu\nu}$ because factors of $\sqrt{\beta}$ can cancel in the variation of the action, i.e., we have to take the limit only after having performed the variation with respect to $\delta_\zeta f_{\mu\nu}$ and $\delta_\zeta g_{\mu\nu}.$ Eventually, $g_{\mu\nu}$ and $\chi$ become gauge invariant (modulo diffeomorphisms) in the limit $\beta\rightarrow \infty,$ while $f_{\mu\nu}$ enjoys the enhanced scalar gauge symmetry in eq.~\eqref{gauge-transf-lambda2} that kills two degrees of freedom in addition to the other four that are eliminated by the constraint $\nabla^{\mu}f_{\mu\nu}=\nabla_\nu f$.  Thus, the  spin-$2$ field $f_{\mu\nu}$ of mass $m_2^2=\frac{2}{3}\Lambda$ contains $10-4-2=4$ independent degrees of freedom: this confirms its partially massless nature~\cite{Deser:1983mm,Deser:1983tm,Deser:2001us}.

The additional scalar gauge symmetry is inherited by the invariance under local Weyl transformation of the action $S_{\rm W},$ as first noticed in the context of cosmological conformal gravity~\cite{Deser:2012euu}.\footnote{The authors in~\cite{Deser:2012euu} were not interested in the theory~\eqref{action-stelle} but only in the cosmological version of $S_{\rm W}$ with $\Lambda>0.$ In this scenario the degrees of freedom are six: four helicities from the partially massless graviton and two helicities from the massless graviton (one of the two fields must be ghost-like).  The helicity-zero is absent because the massive spin-$2$ field has mass equal to $\frac{2}{3}\Lambda$ by default, i.e., no partially massless limit needs to be taken in~\cite{Deser:2012euu}.} Using the field equations of $f_{\mu\nu},$ i.e., $f_{\mu\nu}=\sqrt{\frac{3\beta}{2\Lambda}}(R_{\mu\nu}-\frac{1}{6}g_{\mu\nu}R-\frac{1}{3}\Lambda g_{\mu\nu}),$ the action~\eqref{action-spin-2-lambda-5} can be recast as $S_{\rm W}[g]+S_{00}[g,\phi,\chi]$. Therefore, in the intermediate step we are considering the spin-$2$ sector is described by cosmological conformal gravity, while the spin-$0$ sector by $S_{\rm 00}.$ Our result is in agreement with a more general theorem~\cite{Joung:2014aba} according to which a self-interacting partially massless graviton can be coupled to the metric if the latter also transforms under the gauge symmetry and, in addition, if one of the two gravitons is ghost-like. Below we show that, after diagonalizing the kinetic operator for $g_{\mu\nu}$ and $f_{\mu\nu}$ and expanding in field fluctuations, the limit $\beta\rightarrow \infty$ actually kills all the interactions mediated by the spin-$2$ sector.

Comparing eqs.~\eqref{S00-lambdaneq0} and~\eqref{S00-lambda=0} we can notice that when $\Lambda> 0$ the helicity-zero $\chi$ acquires a mass. In particular,
if we expand around the minimum $\chi=0$,  we find that the helicity-zero not only is a ghost but also a tachyon:
\begin{eqnarray}
	S_{00}\supset\int{\rm d}^4x\sqrt{-g}\left[\frac{1}{2}\nabla_\mu\chi\nabla^\mu\chi+\frac{M_0^2}{2}\chi^2+\cdots\right]\,,
	\label{chi-expans-min}
\end{eqnarray}
where its negative mass squared is given by
\begin{eqnarray}
M_0^2\equiv -\frac{4}{3}\Lambda \,.
	\label{mass-helicity-zero}
\end{eqnarray}
This feature is in agreement with the expectations from group representation theory in de Sitter spacetime according to which in the limit $m_2^2\rightarrow \frac{2}{3}\Lambda$ a massive spin-$2$ field propagating in the de Sitter background  splits into a partially massless graviton of mass squared $\frac{2}{3}\Lambda$ and a massive scalar field of mass squared $-\frac{4}{3}\Lambda$~\cite{Evans-group-de-sitter,Fronsdal-group-de-sitter,DeRham:2018axr}. However, it is important to emphasize that our result is completely non-linear and, for the time being, no particular background has been specified. Moreover, since $\chi$ is ghost-like its self-potential is bounded from below despite being a tachyon. 

In summary, the particle spectrum in the partially massless limit contains a massless spin-$2$ graviton with two helicities ($\pm 2$), a ghost-like partially massless graviton with four helicities ($\pm 2$ and $\pm 1$), a ghost-like tachyonic scalar field (corresponding to the helicity-zero), and a normal scalar field coming from the $R^2$ term: in total we have $2+4+1+1=8$ degrees of freedom.

\subsubsection{Compatible backgrounds}\label{sec:comp-backg}

We are now going to investigate which backgrounds $(\bar{g}_{\mu\nu},\,\bar{f}_{\mu\nu},\,\bar{\phi},\,\bar{\chi})$ are allowed. We define the vacua to be covariantly constant, i.e., $\bar{\nabla}_\rho \bar{f}_{\mu\nu}=0,$ $\bar{\nabla}_\rho\bar{\phi}=\partial_\rho\bar{\phi}=0$ and  $\bar{\nabla}_\rho\bar{\chi}=\partial_\rho\bar{\chi}=0.$ Replacing metric and fields with their backgrounds in the field equations of $f_{\mu\nu}$ and acting with a covariant derivative we obtain $\bar{\nabla}_\rho \bar{G}_{\mu\nu}=0.$
If we contract $\rho$ and $\mu$ we have $\bar{\nabla}_\nu \bar{R}=0,$ thus substituting back we find that the metric background has constant Ricci tensor:
\begin{equation}
\bar{\nabla}_\rho\bar{R}_{\mu\nu}=0\qquad \Rightarrow \qquad \bar{R}_{\mu\nu}=\frac{\bar{R}}{4}\bar{g}_{\mu\nu}\,.\label{einstein-space-cond}
\end{equation}
Using again the field equations of $f_{\mu\nu}$ we find that the background $\bar{f}_{\mu\nu}$ is conformally related to $\bar{g}_{\mu\nu}$ as
\begin{equation}
\bar{f}_{\mu\nu}=\frac{\bar{f}}{4}\bar{g}_{\mu\nu}\,,\qquad \bar{f}=\frac{1}{3}\frac{\tilde{M}_{\rm p}}{m_2^2}(\bar{R}-4\Lambda)\,,\qquad \bar{\nabla}_\rho\bar{f}=\partial_\rho\bar{f}=0\,.\label{conformal-relat-back}
\end{equation}
From the shape of the potential of the scalar fields $\phi$ and $\chi$ in eq.~\eqref{S00-lambdaneq0} we understand that the vacuum is given by $\bar{\phi}=0=\bar{\chi}$ which also implies $\bar{\omega}=0$ (see eq.~\eqref{tilde-phi-def}). Since $\omega$ is proportional to $R-4\Lambda$ by construction, it follows that the background metric must satisfy $\bar{R}=4\Lambda,$ and using~\eqref{conformal-relat-back} we also get $\bar{f}=0.$  Therefore, the compatible metric background and field vacua are
\begin{equation}
	\bar{g}_{\mu\nu}\!:\, \bar{R}_{\mu\nu}=\Lambda\bar{g}_{\mu\nu}\,,\qquad \bar{f}_{\mu\nu}=0\,,\qquad \bar{\phi}=0\,,\qquad \bar{\chi}=0\,.\label{compat-backgr-g-varphi}
\end{equation}

This means that when $\Lambda \neq 0$ only Einstein spacetimes -- such as de Sitter and Kerr-de Sitter black holes -- are allowed as a background metric. Note that in the case $\Lambda=0$ discussed in sec.~\ref{sec:beta-inf-lambda=0}, the allowed metric backgrounds would be Ricci flat, i.e., $\bar{R}_{\mu\nu}=0.$

\subsubsection{Decoupling of the spin-$2$ sector}\label{secsub:decoup}

As already mentioned above, strictly speaking, the expression of the covariant action~\eqref{action-spin-2-lambda-5} is valid for large but still finite $\beta$, i.e., up to corrections of order $\mathcal{O}(1/\sqrt{\beta})$. To fully take the limit $\beta \rightarrow \infty$ and thus make the dependence on $\beta$ disappear from the final result we now diagonalize the kinetic term for $g_{\mu\nu}$ and $f_{\mu\nu}$ and expand in field fluctuations. In this way we can identify the interaction couplings and understand whether some of the degrees of freedom decouple. Remarkably, unlike the case $\Lambda=0,$ we will find that in the partially massless limit the spin-$2$ sector becomes free and completely decouples. 

Performing the transformation $g_{\mu\nu}\rightarrow g_{\mu\nu}-2f_{\mu\nu}/\sqrt{2\Lambda\beta/3}$ and expanding in the field fluctuations $f_{\mu\nu}$ around the background $\bar{f}_{\mu\nu}=0$ we can diagonalize the kinetic operator and write
\begin{eqnarray}
	S[g,f,\chi,\phi] = \frac{2\Lambda \beta}{3M_{\rm p}^2}S_{\rm EH}[g]+S_{2}[g,f]\big|_{m_2^2=\frac{2}{3}\Lambda}+S_{00}[g-2	f/\sqrt{2\Lambda\beta/3},\phi,\chi]\,,
	\label{action-lambda-diag}
\end{eqnarray}
where $S_{2}[g,f]\big|_{m^2_2=\frac{2}{3}\Lambda}$ is equal to the expression in eq.~\eqref{action-spin-2-2} with $m_2^2=\frac{2}{3}\Lambda.$ 

Next, we expand in metric fluctuations, i.e., $g_{\mu\nu}=\bar{g}_{\mu\nu}+2h_{\mu\nu}/\sqrt{2\Lambda\beta/3},$ where we now know that $\bar{g}_{\mu\nu}$ must be of Einstein type.
Unlike the case $\Lambda=0$ analysed in sec.~\ref{sec:beta-inf-lambda=0},  in the presence of a cosmological constant the interactions induced by $h_{\mu\nu}$ and $f_{\mu\nu}$ scale as powers of $1/\tilde{M}_{\rm p}\sim 1/\sqrt{\Lambda\beta}$ instead of $1/M_{\rm p}$ (see also sec.~\ref{sec:int-coup}). This means that in the limit $\beta\rightarrow \infty$ all interactions involving the two gravitons will vanish. The final resulting action in the  partially massless limit reads
\begin{eqnarray}
	S[\bar{g},h,f,\chi,\phi] \!&\!=\!&\! S^{(2)}_{\rm EH}[\bar{g},h]-S_{\rm FP}[\bar{g},f]\big|_{m_2^2=\frac{2}{3}\Lambda}- \int {\rm d}^4x\sqrt{-\bar{g}}\,\Lambda\left(f_{\mu\nu}f^{\mu\nu}-\frac{1}{2}f^2\right)\nonumber\\[2mm]	
	\!&&\!+S_{00}[\bar{g},\phi,\chi]\,,
	\label{action-lambda-decoupling}
\end{eqnarray}
where we have used $\bar{R}_{\mu\nu}=\Lambda \bar{g}_{\mu\nu}$; the first term $S^{(2)}_{\rm EH}$ denotes the Einstein-Hilbert action up to quadratic order in $h_{\mu\nu}$. 
Therefore,  the limit $\beta\rightarrow \infty$ with $\Lambda> 0$  kills all the interactions mediated by the massless and partially massless gravitons ($2+4=6$ degrees of freedom), while the two scalars survive with their self and mutual interactions. This discriminating feature between the two sets  of fields is due to the fact that the interactions mediated by the spin-$0$ sector scale as powers of $1/\bar{M}_{\rm p}$ which is independent of $\beta.$

A comment is in order. As discussed around eq.~\eqref{action-spin-2-lambda-5}, at the non-linear level we can covariantly identify $f_{\mu\nu}$ as a partially massless graviton only if both gravitons transform according to~\eqref{gauge-transf-lambda2}, which is in agreement with a more general theorem~\cite{Joung:2014aba}. However, strictly speaking, in our case the exact identification of a partially massless graviton can only be made if we fully take the limit $\beta\rightarrow \infty$ in which both massless and partially massless gravitons become free. This means that $g_{\mu\nu}$ becomes invariant under~\eqref{gauge-transf-lambda2} and the quadratic action for $f_{\mu\nu}$ is invariant under $\delta_\zeta f_{\mu\nu}$ alone. 

In summary, in the limit $\beta\rightarrow \infty$ with $\Lambda>0$ the theory of Quadratic Gravity in eq.~\eqref{action-stelle} gets surprisingly simpler and the resulting dynamics is solely determined by the action $S_{00}[\bar{g},\phi,\chi]$ that describes two interacting scalar fields (one is ghost-like and tachyonic) living in an Einstein metric background. 
It is important to emphasize that the phenomenon of ``asymptotic freedom'' of the spin-$2$ sector in the presence of a positive cosmological constant  drastically distinguishes the scenario $\Lambda >0$ from the one with $\Lambda=0$ where, instead, the spin-$2$ interactions survive.

\subsubsection{Interaction with matter}\label{sec:int-matt}

Given a generic matter field $\Psi,$ the action~\eqref{action-spin-2-lambda-4} is extended to
\begin{eqnarray}
	S^\prime[g,f,\chi,\phi; \Psi]=S[g,f,\chi,\phi]+S_{\rm m}\Big[e^{\sqrt{\frac{2}{3}}(\phi-\chi)/\bar{M}_{\rm p}}g,\Psi\Big]\,,
	\label{action-spin-2-lambda-5-matter}
\end{eqnarray}
where $S_{\rm m}$ denotes the matter action. The total action is invariant under the gauge transformations~\eqref{gauge-transf-lambda-2}. 

After diagonalizing the kinetic operator for $g_{\mu\nu}$ and $f_{\mu\nu},$ and expanding in the gravitons fluctuations, the matter action formally becomes
\begin{equation}
S_{\rm m}\Big[e^{\sqrt{\frac{2}{3}}(\phi-\chi)/\bar{M}_{\rm p}}\left(\bar{g}+2(h-f)/\tilde{M}_{\rm p}\right),\Psi\Big]\,,
\end{equation}
where $\bar{g}_{\mu\nu}$ is still of Einstein type since the proof that selects~\eqref{compat-backgr-g-varphi} as the only compatible backgrounds still holds by evaluating $\Psi$ at its vacuum value that minimizes the potential.

Since in the limit $\beta\rightarrow \infty$ we have  $1/\tilde{M}_{\rm p}\rightarrow 0$, the interactions between matter and spin-$2$ gravitational sector also vanish. Whereas, the interactions between  matter and  spin-$0$ sector are governed by the coupling $1/\bar{M}_{\rm p}$ which remains finite. Therefore, the resulting gravity-matter action reads
\begin{eqnarray}
	S^\prime[\bar{g},h,f,\chi,\phi;\Psi] &=& S^{(2)}_{\rm EH}[\bar{g},h]-S_{\rm FP}[\bar{g},f]\big|_{m_2^2=\frac{2}{3}\Lambda}- \int {\rm d}^4x\sqrt{-\bar{g}}\,\Lambda\left(f_{\mu\nu}f^{\mu\nu}-\frac{1}{2}f^2\right)\nonumber\\[2mm]	
	&&+S_{00}[\bar{g},\phi,\chi]+S_{\rm m}\Big[e^{\sqrt{\frac{2}{3}}(\phi-\chi)/\bar{M}_{\rm p}}\bar{g},\Psi\Big]\,.
	\label{action-decoupling-matter}
\end{eqnarray}

In the case of a matter action that is invariant under local Weyl transformation the interactions between $(\phi,\chi)$ and $\Psi$ would disappear. In such a scenario gravitational and matter degrees of freedom would be completely decoupled in the limit $\beta\rightarrow \infty$ with $\Lambda >0,$ and the only way in which gravity would be felt by the matter fields is through the presence of a fixed curved metric background $\bar{g}_{\mu\nu}$. A more detailed analysis of the matter interaction in Quadratic Gravity in the limit $\beta\rightarrow \infty$ will be part of a future work.

Finally, note that when $\Lambda=0$ the interaction couplings are proportional to powers of $1/M_{\rm p}$ as shown in sec.~\ref{sec:beta-inf-lambda=0}. Therefore, in that case the interactions between the two massless gravitons and the matter fields do not vanish.

\section{Discussion}\label{sec:phys-impl}

So far we have studied the limit $\beta\rightarrow \infty$ of the gravitational action~\eqref{action-stelle} as a classical question. Quantum mechanically this limit has a  precise physical meaning since the quadratic-curvature coefficients run with the energy and, in particular, $\beta$ tends to infinity at high energies. The aim of this section is to argue that the results obtained in sec.~\ref{sec:decouplings} can be relevant for the high-energy limit of Quadratic Gravity. Unlike the previous ones, this section has some degree of speculation and will leave room for future investigations.

\subsection{Partial masslessness as high-energy limit}

Let us define the couplings
\begin{equation}
g_0^2\equiv \frac{2}{\alpha}\,,\qquad g_2^2\equiv \frac{2}{\beta}\,.
\label{coupl-g02}
\end{equation}
Using methods of local perturbative QFT one can derive the following 1-loop renormalization group equations (RGEs) for the couplings in the renormalized effective action~\cite{Julve:1978xn,Fradkin:1981iu,Avramidi:1985ki,Salvio:2014soa,Anselmi:2018ibi}:\footnote{Note that the RGEs in~\eqref{RGE-eqs} have been derived by looking at the divergent part of the $1$-loop quantum effective action. As mentioned in Ref.~\cite{Buccio:2023lzo}, it would be interesting to re-derive them by looking at renormalized physical quantities -- such as scattering amplitudes and cross sections at $1$-loop -- and check whether the results match.}
\begin{equation}
\frac{{\rm d}g_0^2}{{\rm d}\tau}=\frac{5}{3}g_2^4+5g_2^2g_0^2+\frac{5}{6}g_0^4\,,\qquad \frac{{\rm d}g_2^2}{{\rm d}\tau}=-\frac{133}{10}g_2^4\,,
\label{RGE-eqs}
\end{equation}
where the modified minimal scheme is assumed, $\tau=\frac{1}{(4\pi)^2}\log (\mu/\mu_0)$ contains the dependence on the energy $\mu,$ and $\mu_0$ is some reference energy scale. 

Formally analogous differential equations can also be derived for $M_{\rm p}^2$ and $\Lambda$~\cite{Julve:1978xn,Fradkin:1981iu,Avramidi:1985ki,Salvio:2014soa,Anselmi:2018ibi}. However, their running is not physical~\cite{Anber:2011ut,Donoghue:2019clr,Donoghue:2022eay}: once we fix the renormalized values of the Planck mass and the cosmological constant they will not change by varying the energy.\footnote{A way to understand this fact is to ask what kind of finite momentum-dependent form factors can be introduced into the quantum effective action to capture the running of $M_{\rm p}$ and $\Lambda$. The only way is to write $\int {\rm d}^4x\sqrt{-g}[F(\Box)R-G(\Box)2\Lambda],$ where $G(\Box)$ and $F(\Box)$ are some diffeomorphism invariant differential operators. However, metric compatibility implies that the $\Box$-dependent contributions are total derivatives that can be discarded.\label{foot-lambda}}

The solutions of the RGEs~\eqref{RGE-eqs} are given by~\cite{Avramidi:1986mj}
\begin{equation}
g_0^2(\tau)=\frac{c_+ g_2^{2p}(\tau)-c_- g_*^{2p}}{g_2^{2p}(\tau)-g_*^{2p}}g_2^2(\tau)  \,,\qquad g_2^2(\tau)=\frac{g_2^2(0)}{1+\frac{133}{10}g_2^2(0)\tau}\,, \label{sol-g0-2}
\end{equation}
where we have defined
\begin{equation}
	 p\equiv \frac{\sqrt{6049}}{57}\,,\qquad c_\pm\equiv\frac{1}{50}(-549\pm 7\sqrt{6049})<0\,,\qquad g_*^{2p}\equiv\frac{g_0^2(0)-c_+g_2^2(0)}{g_0^2(0)-c_-g_2^2(0)}g_2^{2p}(0)\,,
	\label{sol-g0}
\end{equation}
and $g_i^2(0)=g_i^2(\tau=0)=g_i^2(\mu=\mu_0),$ $i=0,2.$

Since we are working with the positive signs $\alpha>0$ and $\beta>0$ (see sec.~\ref{sec:action}), we have  $g_2^2(0)>0$ and $g_0^2(0)>0,$ which also imply $g_*^{2p}>0,$ thus it follows that the coupling $g_2^2(\mu)$ goes to zero in the high-energy limit, while $g_0^2(\mu)$ grows with the energy. In other words, at high energies the Weyl-squared coefficient $\beta(\mu)$ grows and tends to infinity in the limit $\mu\rightarrow \infty$, while $\alpha(\mu)$ decreases. 
The coupling $g_0^2$ becomes strong at high-energies and a non-perturbative treatment may be required. It is worth mentioning that in the context of adimensional gravity (Agravity~\cite{Salvio:2014soa}), where the action is scale invariant and contains only the quadratic-curvature terms, it was claimed that $g_0^2$ grows to infinity without hitting a Landau pole~\cite{Salvio:2017qkx}. 
As we will further explain below, what is important in our case is that $\beta$ tends to infinity in the high-energy limit without breaking perturbativity.

We now argue that the analysis in the previous section can indeed be useful for understanding the high-energy behavior of Quadratic Gravity, in particular when the cosmological constant is non-zero and positive.  In general, the full renormalized quantum effective action, which depends on the running couplings $g_0^2(\mu)$ and $g_2^{2}(\mu),$ will also contain finite quantum corrections which were not taken into account in sec.~\ref{sec:decouplings} and that, in principle, could affect our results. However, the following two arguments suggest that   quantum corrections can be reliably neglected in the high-energy limit at least when $\Lambda> 0.$
\begin{enumerate}
	
	\item The interactions mediated by the spin-$2$ sector remain weak and completely vanish in the  limit $g_2^2\rightarrow 0$ in such a way that the massless graviton and the partially massless graviton become free. Since the loop expansion is controlled by the coupling $g_2^2,$ all the quantum corrections in the spin-$2$ sector can be neglected in the deep UV. This is somehow similar to what happens in Yang-Mills theory where the gluons are weakly interacting at high energy and can be described as classical waves plus small quantum corrections. 
	
	\item Since the coupling $g_0^2$ grows at high energy, the action $S_{00}$ becomes large. This suggests that the path integral is dominated by the classical configurations that minimize the action. Thus, quantum corrections in the spin-$0$ sector can be neglected at the expense of a non-perturbative treatment. The behavior of $\alpha= 1/2g_0^2$ is irrelevant for the spin-$2$ sector:  even if $\alpha\rightarrow 0$ (i.e., $g_0^2\rightarrow \infty$) the result of the limit $\beta\rightarrow \infty$ is unchanged. 
	
\end{enumerate}

It is important to point out that the first argument above does not apply when $\Lambda=0$ because in this case the interaction terms remain finite and scale as powers of $1/M_{\rm p}.$  We are able to justify the use of the classical action in the high-energy limit only when the cosmological constant is non-zero and positive.

Therefore, in the light of the results obtained in sec.~\ref{sec:beta-inf-lambdaneq0} we argue that the partial masslessness of the spin-$2$ ghost determines several aspects of Quadratic Gravity in the deep UV.
\begin{itemize}
	
	\item The physical spectrum is organized in a unique way and contains a massless graviton ($\pm 2$), a ghost-like partially massless graviton ($\pm 2, \pm 1$), a ghost-like tachyonic helicity-zero, and a normal massive spin-$0$.
	
	\item The resulting actions~\eqref{action-spin-2-lambda-5} or~\eqref{action-lambda-decoupling} enjoy an enhanced scalar gauge symmetry given by~\eqref{gauge-transf-lambda2} which underlies the reason why the massive spin-$2$ splits into a partially massless graviton plus a tachyonic scalar field.
	
	\item The spin-$2$ sector becomes free and completely decouples. The only relevant contribution to the high-energy dynamics of Quadratic Gravity is given by the action $S_{00}[\bar{g},\phi,\chi]$ in eq.~\eqref{S00-lambdaneq0}. An interesting feature to emphasize is that, although the coupling $g_2$ vanishes in the high-energy limit regardless of the value of the cosmological constant, this phenomenon of ``asymptotic freedom'' of the spin sector-$2$ is valid only when $ \Lambda>0$.\footnote{It would be interesting to check if this property of ``asymptotic freedom'' of the spin-$2$ couplings when $\Lambda>0$  manifests in the high-energy/short-distance behavior of physical observables, e.g., correlators at $1$-loop.}
	
\end{itemize}

We can think of the early Universe (even pre-inflationary epoch) as an arena where the high-energy limit of Quadratic Gravity could play a relevant role in describing physics. Our analysis might indicate that at very early times (i.e., very high temperature and so high energy) the only dynamical degrees of freedom are the two scalar fields $\phi$ and $\chi$ living in a fixed Einstein-type spacetime, e.g., a maximally symmetric one such as de Sitter, and the interactions of all other fields, including gravitons, are switched off. Then, as the temperature starts dropping down all other interactions switch on and both the graviton and spin-$2$ ghost become active interacting degrees of freedom. However, we expect such a process to be discontinuous because the two particle spectra when $\beta=\infty$ or $\beta\neq \infty$ are drastically different. One way to interpret this physical scenario might be in terms of a \textit{phase transition} and view the coupling $g_2^2=2/\beta$ as an \textit{order parameter}: when $g_2^2=0$ we are in a phase that is determined by the partially massless symmetry and in which there are two dynamical and interacting scalar degrees of freedom, a non-interacting massless graviton and a non-interacting partially massless ghost-like graviton; when $g_2^2\neq 0$ we end up in a different phase containing a massless graviton, a massive spin-$2$ ghost and a normal massive scalar field, all interacting. A more concrete and rigorous investigation of this idea will be part of future work. 

\paragraph{Remark.} Strictly speaking, the more precise statement to make is that the partially massless limit is related to the infinite-energy limit of Quadratic Gravity, otherwise for large but finite values of $\mu$ we would have $\beta\neq \infty$ and $m_2^2\neq \frac{2}{3}\Lambda.$ However, for large but finite $\beta$ we can consider the action~\eqref{action-spin-2-lambda-5} plus corrections of order $\mathcal{O}(f/\sqrt{\beta})$ that act as small explicit symmetry breaking terms. Furthermore, as a future investigation it would also be interesting to understand if higher-loop corrections and/or non-perturbative effects can provide solutions to the RGEs such that $g_2^2\rightarrow 0$ for finite $\mu$ so that partial masslessness would manifest at large but finite energies. 

\subsection{Remarks on the action $S_{00}$} 

Some of the interactions terms in $S_{00}$ are proportional to $m_0^2=M^2_{\rm p}/\alpha$. This means that if $\alpha$ continues to decrease as the energy increases, strong coupling can arise and a non-perturbative treatment is required. In what follows we  make some remarks on $S_{\rm 00}$.

It is convenient to recast $S_{00}$ in canonical form. This can be done by performing the following field transformations: 
\begin{equation}
\phi=-\sqrt{\frac{3}{2}}\bar{M}_{\rm p}\log\left(\frac{\tilde{\chi}+\tilde{\phi}}{\tilde{\chi}-\tilde{\phi}}\right)\,,\qquad \chi=-\sqrt{\frac{3}{2}}\bar{M}_{\rm p}\log\left(\frac{\tilde{\chi}^2-\tilde{\phi}^2}{6\bar{M}_{\rm p}^2} \right)\,,\label{canon-phichi}
\end{equation}
through which the action becomes
\begin{equation}
S_{00}[\bar{g},\tilde{\phi},\tilde{\chi}]=\int {\rm d}^4 x\sqrt{-\bar{g}}\left[-\frac{1}{2}\partial_\mu\tilde{\phi}\partial^\mu\tilde{\phi} +\frac{1}{2}\partial_\mu\tilde{\chi}\partial^\mu\tilde{\chi}-V\left(\tilde{\phi},\tilde{\chi}\right)\right]\,,
\end{equation}
where the potential
\begin{equation}
V\left(\tilde{\phi},\tilde{\chi}\right)=\frac{\Lambda}{36\bar{M}^2_{\rm p}}\left(\tilde{\chi}^2-\tilde{\phi}^2-6\bar{M}_{\rm p}^2 \right)^2 + \frac{m_0^2}{12\bar{M}^2_{\rm p}}\tilde{\phi}^2\left(\tilde{\chi}+\tilde{\phi}\right)^2
\end{equation}
is always non-negative as $\Lambda>0$ and $m_0^2>0;$ although $\tilde{\chi}$ is tachyonic, the ghost-like nature guarantees that its self-potential is positive definite. Note that when $\Lambda =0$ we recover  the action obtained in~\cite{Hinterbichler:2015soa}.

Due to the logarithms in~\eqref{canon-phichi}, the region in field space covered by $(\chi,\phi)$ corresponds to $\tilde{\chi}^2>\tilde{\phi}^2,$ $\tilde{\chi}>0.$ The action is invariant under the transformation $(\tilde{\phi},\tilde{\chi})\rightarrow (-\tilde{\phi},-\tilde{\chi}).$
The minimum $(\phi=0,\,\chi=0)$ is mapped into $(\tilde{\phi}=0,\,\tilde{\chi}=\sqrt{6}\bar{M}_{\rm p})$. Moreover, in the field space of the new variables we also have a new stationary point which is a saddle given by $(\tilde{\phi}=0,\,\tilde{\chi}=0);$ it disappears when $\Lambda=0.$

As a simple exercise let us compute the Hamiltonian density in a de Sitter background metric that in flat FLRW coordinates reads
\begin{equation}
{\rm d}s^2=-{\rm d}t^2+a^2(t){\rm d}\vec{x}^2\,,\qquad \sqrt{-\bar{g}}=a^3(t)\,,\qquad a(t)=e^{\sqrt{\frac{\Lambda}{3}}t}\,.
\end{equation}
It is easy to show that the Hamiltonian density is given by
\begin{equation}
\mathcal{H}=a^3(t)\left[\frac{1}{2}(\partial_0\tilde{\phi})^2-\frac{1}{2}(\partial_0\tilde{\chi})^2+\frac{1}{2a^2(t)}\left(|\vec{\nabla}\tilde{\phi}|^2-|\vec{\nabla}\tilde{\chi}|^2\right)+V\left(\tilde{\phi},\tilde{\chi}\right) \right]\,.
\end{equation}

For a general field configuration the Hamiltonian can in principle become negative due to the contribution associated with the ghost-like helicity-zero. However, we believe that more work is required before making any claim about the presence of an instability. First of all, the action $S_{00}$ does not have a higher-derivative form, therefore we should not expect the Ostrogradsky theorem~\cite{Woodard:2015zca} to apply and, in addition, runaway solutions need not appear. It is also worth mentioning that recent studies~\cite{Deffayet:2021nnt,Deffayet:2023wdg} showed that unbounded Hamiltonians with ghosts can still be stable given certain types of scalar potentials. It would be very interesting to solve non-perturbatively and at a full non-linear level the field equations for $\phi$ and $\chi$ and determine their dynamics. A detailed analysis of $S_{00}$ and its (in)stability will be part of a separate work.

\section{Summary and conclusions}\label{sec:conclus}

In this paper we have studied the limit $\beta\rightarrow \infty$ of the Quadratic Gravity action in eq.~\eqref{action-stelle} and found several interesting results.  To perform the limit we recast the action in a second-order canonical form (in sec.~\ref{sec:action}) and kept the canonically normalized fields fixed.  In sec.~\ref{sec:decouplings} we asked whether the limit is regular and what is the resulting theory; the main findings can be summarized as follows.
\begin{itemize}
	
	\item The limit $\beta\rightarrow \infty$  crucially depends on the presence of the cosmological constant: when $\Lambda=0$ it corresponds to a massless limit $m_2^2\rightarrow 0$ for the spin-$2$ massive ghost; whereas, when $\Lambda>0$ we obtain a partially massless limit $m_2^2\rightarrow \frac{2}{3}\Lambda.$ Moreover, a non-zero cosmological constant affects the interaction couplings in a non-trivial way.
	
	\item These types of limits are known to give rise to strong couplings in standard ghost-free theories of massive gravity~\cite{Hinterbichler:2011tt,deRham:2014zqa,DeRham:2018axr}. However, here we have shown that the renormalizability property of Quadratic Gravity (valid in $D=4$) and the ghost-like nature of the massive spin-$2$ field together guarantee that both limits are regular. 
	
	\item The resulting theories in the two cases $\Lambda=0$ and $\Lambda>0$ turn out to be very different. In the former scenario, the spin-$2$ ghost splits into five ghost-like degrees of freedom: a massless graviton ($\pm 2$ helicities), a massless photon ($\pm 1$ helicities) and a massless scalar (helicity-zero). 
	Whereas, in the latter scenario the spin-$2$ ghost splits into a ghost-like partially massless graviton ($\pm 2,$ $\pm 1$ helicities) of mass $m_2^2=\frac{2}{3}\Lambda$ and a ghost-like tachyonic scalar (helicity-zero) of mass $M_0^2=-\frac{4}{3}\Lambda.$ 
	The complete particle spectrum of Quadratic Gravity also contains the normal massless graviton coming from the Einstein-Hilbert action and the massive scalar coming from the $R^2$ term. In total the degrees of freedom are always eight.
	
	\item  In sec.~\ref{secsub:decoup},  we found that in the limit $\beta\rightarrow \infty$  the interactions mediated by the spin-$2$ sector (massless graviton $+$ partially massless graviton) vanish, while the spin-$0$ sector (normal massive scalar $+$ ghost-like tachyonic scalar) remains interacting and is described by the action~\eqref{S00-lambdaneq0}. This difference between the two sets of fields is due to the fact that the spin-$2$ sector interacts with couplings proportional to powers of $1/\tilde{M}_{\rm p}=(M_{\rm p}^2+\frac{2}{3}(2\alpha+\beta)\Lambda)^{-1},$ while the spin-$0$ sector with powers of $1/\bar{M}_{\rm p}=(M_{\rm p}^2+\frac{4}{3}\alpha\Lambda)^{-1}$. No decoupling occurs when $\Lambda=0$ because the interaction couplings are proportional to powers of $1/M_{\rm p}$ in both sectors. 
	
	\item We showed that in the limit $\beta\rightarrow \infty$ new enhanced gauge symmetries emerge. When $\Lambda=0,$ the ghost-like massless spin-$2$ and spin-$1$ exhibit the gauge symmetries $\delta_\xi f_{\mu\nu}=\nabla_\mu\xi_\nu+\nabla_\nu\xi_\mu$ and $\delta_\xi A_\mu= \nabla_\mu \xi,$ respectively. When $\Lambda>0$ the partially massless graviton exhibits the scalar gauge symmetry $\delta_\zeta f_{\mu\nu}=\nabla_\mu\nabla_\nu\zeta+\frac{\Lambda}{3}g_{\mu\nu}\zeta.$ 

	\item In sec.~\ref{sec:comp-backg} we explained that the compatible metric backgrounds are Ricci flat ($\bar{R}_{\mu\nu}=0$) when $\Lambda=0$ and Einstein ($\bar{R}_{\mu\nu}=\Lambda \bar{g}_{\mu\nu}$) when $\Lambda >0.$ The backgrounds, or in other words the vacua, of the additional fields are $\bar{f}_{\mu\nu}=0$, $\bar{\phi}=0$ and $\bar{\chi}=0.$

\end{itemize}

Subsequently, in sec.~\ref{sec:phys-impl} we argued that our analysis of the limit $\beta\rightarrow \infty$ can be relevant for the dynamics of Quadratic Gravity in the deep UV regime. Indeed, quantum mechanically the Weyl-squared coefficient $\beta$  runs with the physical energy  and tends to infinity in the infinite-energy limit. In particular, in the case of a non-zero and positive cosmological constant we provided arguments suggesting that quantum corrections can be neglected in the limit of interest and that the discussion in sec.~\ref{sec:beta-inf-lambdaneq0}  literally applies. This means that high-energy aspects of Quadratic Gravity can be understood in terms of the partially massless limit of the spin-$2$ ghost and the dynamics in the infinite-energy limit is fully captured by $S_{00}.$

Since the $R^2$ coefficient $\alpha$ decreases with the energy, some interaction couplings in $S_{00}$ may become strong in such a way that a non-perturbative treatment may be required. We made several remarks on the action $S_{00}$ and argued that a fully non-perturbative and non-linear study is required before making any claim about the (in)stability of the dynamics of the two scalar fields. A more general and deeper analysis will be part of future work.

Let us make some final comments on the role of the cosmological constant. It is important to point out that in all approaches aimed at quantizing the theory of Quadratic Gravity in~\eqref{action-stelle} it is usually assumed that the cosmological constant is negligible~\cite{Salvio:2018crh,Donoghue:2019ecz,Anselmi:2017ygm,Anselmi:2018ibi}. However, contrary to common beliefs, in this work we have shown that there exists a regime (given by $\beta=\infty$) in which the cosmological constant can play a crucial role. In particular, its presence may uniquely determine the high-energy limit  of the theory in terms of a detailed particle spectrum, an enhanced scalar gauge symmetry and a decoupling phenomenon of the spin-$2$ sector. No matter how small $\Lambda$ is, its non-zero and positive value has huge implications.

To our knowledge, this is a unique example where an a priori innocuous cosmological constant can affect high-energy properties in a theory of quantum gravity. In fact, there is an additional interesting aspect to point out: in Quadratic Gravity the cosmological constant can be expressed in terms of other fundamental quantities which are $M_{\rm p}^2,$ $m_2^2,$ $\beta$ and $\alpha$ (or $m_0^2$). This fact simply follows from the expression of the spin-$2$ ghost mass in eq.~\eqref{spin-2-mass} from which we get
\begin{equation}
	\Lambda=\frac{3}{2}\frac{\beta m_2^2-M_{\rm p}^2}{\beta+2\alpha}=\frac{3}{2}m_2^2\frac{\beta-M_{\rm p}^2/m_2^2}{\beta+2M_{\rm p}^2/m_0^2}\,;
	\label{Lambda-formula}
\end{equation}
it is always positive since we are working with $\beta>0,$ $\alpha>0$ and $m_2^2>M_{\rm p}^2/\beta.$ 

Given all these interesting features, we believe that a deeper understanding of the role of the cosmological constant in Quadratic Gravity and, more generally, in quantum gravity is needed. In particular, it would be very interesting to understand if and how the current approaches to the quantization of the spin-$2$ ghost in Quadratic Gravity are influenced by a non-zero cosmological constant.

In conclusion, Quadratic Gravity is a very unique  four-dimensional theory of quantum gravity which is perturbatively renormalizable and that can explain both late and early time cosmological observations, thus describing  physics beyond Einstein's general relativity. The structure of the theory, including its peculiar particle spectrum, is very rich and, as various works have shown in recent years, the more we explore and dig into the theory, the more we discover and learn. We hope that our results will help gain new insights into the stability of the theory at high energy and the role of the cosmological constant in quantum gravity. Therefore, instead of throwing the theory away just because of the unusual spin-$2$ ghost, we strongly believe that Quadratic Gravity as the QFT of the gravitational interaction deserves much more interest and investigation.


\subsection*{Acknowledgements}
I am very grateful to Damiano Anselmi, Marco Piva and especially Fawad Hassan and Kurt Hinterbichler for useful discussions, to Massimo Taronna  for explaining some details of his work~\cite{Joung:2014aba}, and to Tibério de Paula Netto for comments. I would like to thank Elizabeth Colina for her inspiring support and advices.  Nordita is supported in part by NordForsk. 



\bibliographystyle{utphys}
\bibliography{References}


\end{document}